\documentclass[prl,twocolumn,preprintnumbers,amsmath,amssymb,superscriptaddress]{revtex4-1}
\usepackage{graphicx, epsf,bm,amsmath}
\usepackage{subfigure,color}
\newcommand{\br}{{\bf r}}

\newcommand{\bk}{{\bf k}}

\begin{document}
\title{Uncovering the hidden quantum critical point in disordered massless Dirac and Weyl semi-metals}
\author{J. H. Pixley}
\affiliation{Condensed Matter Theory Center and the Joint Quantum Institute, Department of Physics, University of Maryland, College Park, Maryland 20742-4111 USA}
\author{David A. Huse}
\affiliation{Physics Department, Princeton University, Princeton, NJ 08544 USA, and Institute for Advanced Study, Princeton, NJ 08540 USA}
\author{S. Das Sarma}
\affiliation{Condensed Matter Theory Center and the Joint Quantum Institute, Department of Physics, University of Maryland, College Park, Maryland 20742-4111 USA}
\date{\today}

\begin{abstract}
We study the properties of the avoided or hidden quantum critical point (AQCP) in three dimensional Dirac and Weyl semi-metals in the presence of short range potential disorder. By computing the averaged density of states (along with its second and fourth derivative at zero energy) with the kernel polynomial method (KPM) we systematically tune the effective length scale that eventually rounds out the transition and leads to an AQCP.  We show how to determine
the strength of the avoidance, establishing that it is not controlled by the long wavelength component of the disorder.  Instead, the amount of avoidance can be adjusted via the tails of the probability distribution of the local random potentials.  A binary distribution with no tails produces much less avoidance than a Gaussian distribution.
We introduce a double Gaussian distribution to interpolate between these two limits.  As a result we are able to make the length scale of the avoidance sufficiently large so that we can accurately study the properties of the underlying transition (that is eventually rounded out), unambiguously identify its location, and provide accurate estimates of the critical exponents $\nu=1.01\pm0.06$ and $z=1.50\pm0.04$.  We also show that the KPM expansion order introduces an effective length scale that can also round out the transition in the scaling regime near the AQCP.
\end{abstract}

\pacs{71.10.Hf,72.80.Ey,73.43.Nq,72.15.Rn}

\maketitle

\begin{figure}[h!]
\centering
\begin{minipage}{.25\textwidth}
\includegraphics[width=1.1\linewidth]{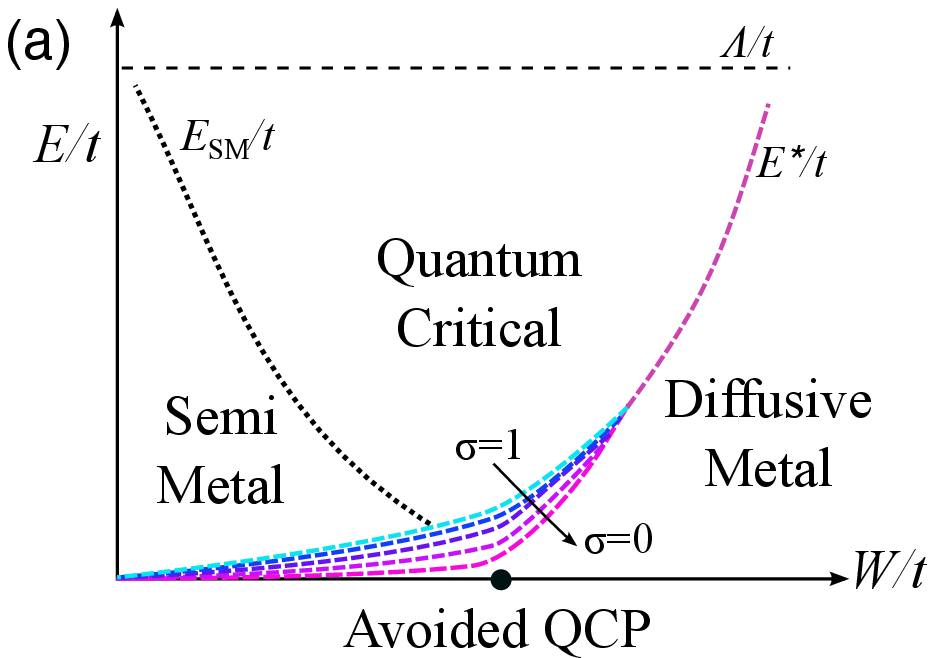}
\end{minipage}
\centering
\begin{minipage}{.25\textwidth}
\includegraphics[width=0.65\linewidth]{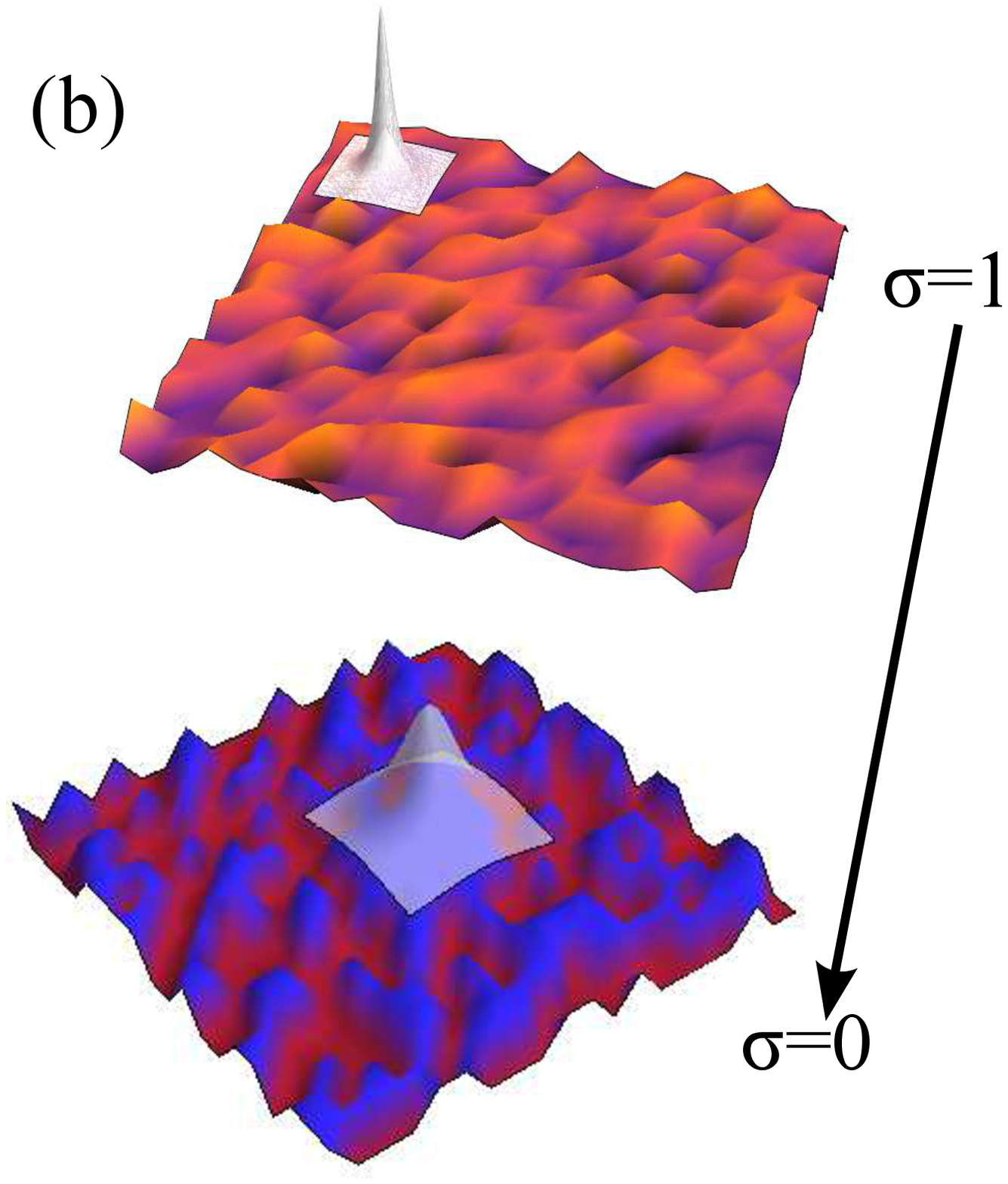}
\end{minipage}
\caption{(color online) (a) Schematic crossover diagram as a function of energy ($E$) and disorder ($W$) with each relevant regime: SM, QC fan, and DM. 
For $E>\Lambda$ the low energy description in terms of a linear dispersion no longer applies and $E_{\mathrm{SM}}$ is set by the distance to the AQCP.
Varying the disorder distribution ($\sigma$) controls the strength of the non-perturbative effects that round out the QCP and (as we will show)
tunes the cross over energy $E^*$ 
 increasing the size of the QC regime.  (b) Schematic of a
disorder profile for a rare configuration and a
rare low-$|E|$ eigenstate that is power law quasi-localized like $\sim1/r^2$ in the DM regime at small $W/t$.  For $\sigma=1$ 
the unbounded tails of the distribution leads to large local fluctuations of the potential on one or two sites that can non-perturbatively produce these rare eigenstates.   
$\sigma\rightarrow0$ 
suppresses the probability to generate 
rare eigenstates, here we expect such a rare state will be produced by a 
large cluster of sites all with the same sign of $W$.}
\label{fig:1}
\end{figure}

Zero temperature quantum phase transitions have become a central pillar to understand various experiments in insulating magnets~\cite{Giamarchi-2008}, two dimensional electron gases~\cite{Sondi-1997}, disordered superconductors~\cite{Weichman-2008}, and strongly correlated electron systems~\cite{Hilbert-2007, Si-2014}. For the case of itinerant quantum phase transitions
the large accumulation of entropy, near the quantum critical point (QCP)  in the finite temperature ``quantum critical fan''~\cite{Sachdev-book} [see Fig.~\ref{fig:1}(a)],
tends to nucleate other broken symmetry phases (e.g. superconductivity~\cite{Broun-2008}) 
that hides the QCP 
and rounds out the critical divergences.
In most cases there are numerous ordering channels and most theories are either biased or have little to no control on whether the transition will become 
avoided.
As a result, identifying a class of models where the avoided QCP is intrinsic to the problem and the avoidance can be studied and controlled in an exact  unbiased fashion is a fundamental question of interest. As we will show, disordered non-interacting Dirac and Weyl semi-metals are a quintessential example.

Recently, there has been a great deal of activity in trying to understand weakly-interacting three dimensional Dirac and Weyl semi-metals.  These materials (such as Cd$_3$As$_2$~\cite{Neupane-2014,Liu-2014,Borisenko-2014}, Na$_3$Bi~\cite{Liu2-2014,Xu-2015}, TaAs~\cite{Xu3-2015,Weng-2015} and NbAs~\cite{Xu2-2015}) have recently been discovered through angle resolved photoemission spectroscopy guided by first principles calculations~\cite{Wan-2011,HWeng-2015,Huang-2015}.  One main theoretical focus has been the effect of short ranged potential disorder~\cite{Fradkin-1986,Goswami-2011,Kobayashi-2014,Brouwer-2014,Bitan-2014,*Bitan-2016,Nandkishore-2014,Pixley-2015,Altland-2015,Sergey-2015,Leo-2015,Sbierski-2015,Pixley2015disorder,Garttner-2015,Liu-2015,Bera-2015,Shapourian-2015,Altland2-2015,Sergey2-2015,Louvet-2016} and the proposed (perturbatively accessible) QCP separating a semi-metal (SM) and diffusive metal (DM) phase which is driven by tuning the strength of disorder. However, due to non-perturbative effects of rare regions~\cite{Nandkishore-2014}, that gives rise to weakly dispersing quasi-localized eigenstates with non-zero level repulsion, this transition is rounded out and becomes an avoided quantum critical point (AQCP)~\cite{Pixley-2016}. As a result, the validity of each previous numerical study of the critical properties of this transition are now called into doubt since these 
did not take into account rare region effects and the hidden character of the QCP. Interestingly, this AQCP is remarkably similar to the QCP becoming hidden via other ordered phases with various numerical studies~\cite{Kobayashi-2014,Brouwer-2014,Pixley-2015,Sbierski-2015,Pixley2015disorder,Garttner-2015,Liu-2015,Bera-2015,Shapourian-2015} observing (at best) only a glimpse of the underlying quantum critical properties. But, in this case there is only one phase, the DM [with a density of states (DOS) $\rho(E)\approx\mathrm{const}$ for $E<E^*$ at zero energy (or temperature) see Fig.~\ref{fig:1}(a)], with cross overs at nonzero energy to the SM regime  ($\rho(E)\sim E^2$ for $E_{\mathrm{SM}}>E>E^*$) and a quantum critical (QC) regime ($\rho(E)\approx|E|$ for $\Lambda>E>E^*$) that is anchored by the AQCP. Since these models are non-interacting, we can study the AQCP in a numerically exact fashion without the complications of
strong correlations. Thus disordered Dirac and Weyl SMs serve as an excellent system to gain fundamental new insights into AQCPs and how zero temperature transitions can show universal scaling before becoming rounded out. 
We show explicitly in the current work how the quantum critical properties 
hidden under the AQCP can be uncovered through numerical simulations by systematically suppressing the non-perturbative (and noncritical) effects. 

In this manuscript we establish how to tune the effective cross-over energy scale ($E^*$) associated with 
non-perturbative effects 
that hide
the QCP, so that we can make $E^*$ small enough at the AQCP to observe the quantum critical scaling regime over a significant energy range [see Fig.~\ref{fig:1}(a)].
There are two properties of the disorder distribution that we separately control and study: (i) the long wave-length component of the disorder, and (ii) the probability to generate a rare eigenstate [see Fig.~\ref{fig:1}(b)].  We find that suppressing the long-wavelength component of Gaussian disorder does not affect the strength of the avoidance, whereas controlling the unbound tails of the disorder distribution can systematically tune $E^*$. Using a binary disorder distribution greatly reduces the probability to generate rare events\cite{Nandkishore-2014}.  This makes the crossover energy scale $E^*$ sufficiently small so that
we can precisely study the 
AQCP and determine accurate estimates of its critical exponents ($\nu=1.01\pm0.06$ and $z=1.50\pm0.04$).  
However, we are never able to completely uncover the AQCP, as it is always rounded out eventually by the effects of rare regions~\cite{Pixley-2016}.

We focus on a three-dimensional tight binding Weyl Hamiltonian in the presence of short range potential disorder, which
 is defined as~\cite{Pixley-2015,Pixley2015disorder,Pixley-2016}
\begin{equation}
H = \sum_{\br, \mu=x,y,z}\frac{1}{2}(it_{\mu}\psi^{\dag}_{\br}\sigma_{\mu}\psi_{\br + \hat{\mu}} + \mathrm{h.c.}) + \sum_{\br} V(\br)\psi^{\dag}_{\br}\psi_{\br}.
\end{equation}
$\psi_{\br}$ is a two component spinor, $\sigma_{\mu}$ are the Pauli operators, and the onsite random disorder potential is $V(\br)$. We consider a cubic lattice of linear size $L$ with twisted periodic boundary conditions on each sample that gives $t_{\mu}=t\exp{(i\theta_{\mu}/L)}$ for a twist $\theta_{\mu}$ in the $\mu$ direction. We consider taking various different choices for the probability distribution for the disorder potential $P[V]$ in order to tune the length scale associated with the AQCP, see Fig.~\ref{fig:1}. We consider five choices for $P[V]$:
a gaussian with zero mean and variance $W^2$, a ``colored'' gaussian with a variance in momentum space $\langle|V(\bk)|^2 \rangle=W^2(\sum_{\mu}\sin(k_{\mu})^2)$, which gives rise to correlated disorder with a vanishing long wavelength component, a box distribution $V(\br) \in [-\tilde{W}/2,\tilde{W}/2]$ (variance $W^2=\tilde{W}^2/12$), a binary distribution which takes values $\pm W$ with equal probability, and a double gaussian distribution that interpolates between the gaussian and binary distributions. For the double gaussian we sample two gaussians with equal probability that have means $\pm W\sqrt{1-\sigma^2}$ and have a standard deviation $W\sigma$, thus the full distribution always has a variance $W^2$. This allows us to 
tune between gaussian and binary distributions, i.e.   
$\sigma\rightarrow 1$ it is a single gaussian and $\sigma\rightarrow 0$ it is the binary distribution.

We use the kernel polynomial method~\cite{Weisse-2006} (KPM) to compute the average DOS
\begin{equation}
\rho(E) = \frac{1}{N_RV}\sum_r^{N_R}\sum_i^{2V}\delta\left(E-E_i(r)\right),
\end{equation}
where $V=L^3$ is the volume, $N_R$ is the number of disorder realizations, and $E_i(r)$ is the $i$th eigenvalue of the $r$th disorder realization.
We average over $N_R=1,000$ disorder realizations that each have a random twist vector $\bm{\theta}=(\theta_x,\theta_y,\theta_z)$ ($\theta_i$ is sampled uniformly between $[0,\pi]$). We take odd $L$ and average over the twist to minimize finite size effects at all $E$~\cite{Pixley-2016}. 
We evaluate the stochastic trace within KPM using normalized random vectors~\cite{Wilson-2016}. 
The KPM expands the DOS in terms of Chebyshev polynomials to an order $N_C$ and we use the Jackson kernel to filter out Gibbs oscillations. 
The Jackson kernel broadens each Dirac-delta function in the DOS into a Gaussian~\cite{Weisse-2006} of width $\pi D/N_C$ (for a bandwidth $D$). As we will show,
this broadening introduces an effective length scale into the problem that is controlled by the expansion order $N_C$, which can also round out the transition when the strength of the avoidance is sufficiently weak (after suppressing non-perturbative effects).

\begin{figure}[h!]
\centering
\begin{minipage}{.25\textwidth}
\includegraphics[width=0.7\linewidth,angle=-90]{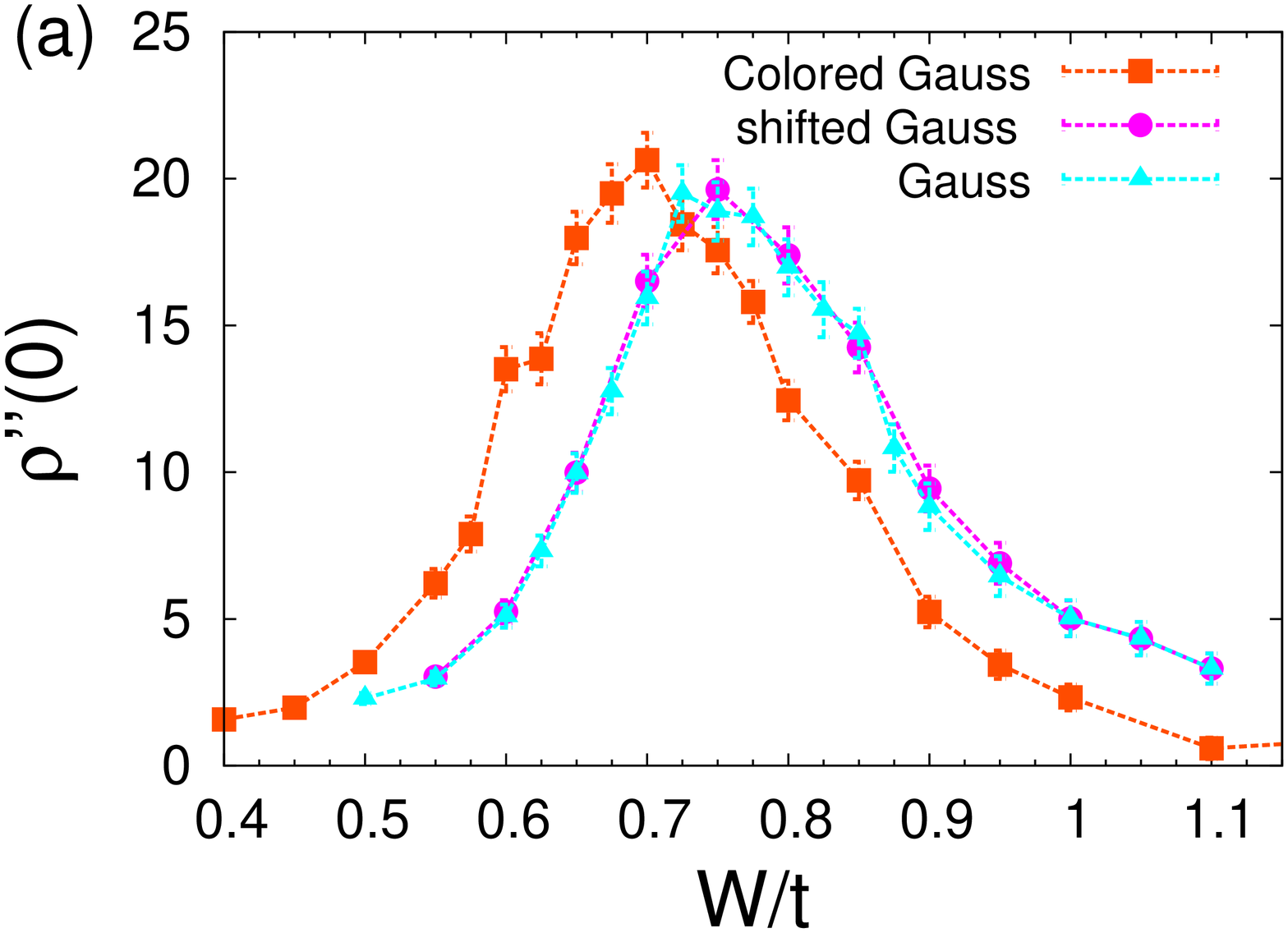}
\end{minipage}\centering
\begin{minipage}{.25\textwidth}
\includegraphics[width=0.7\linewidth,angle=-90]{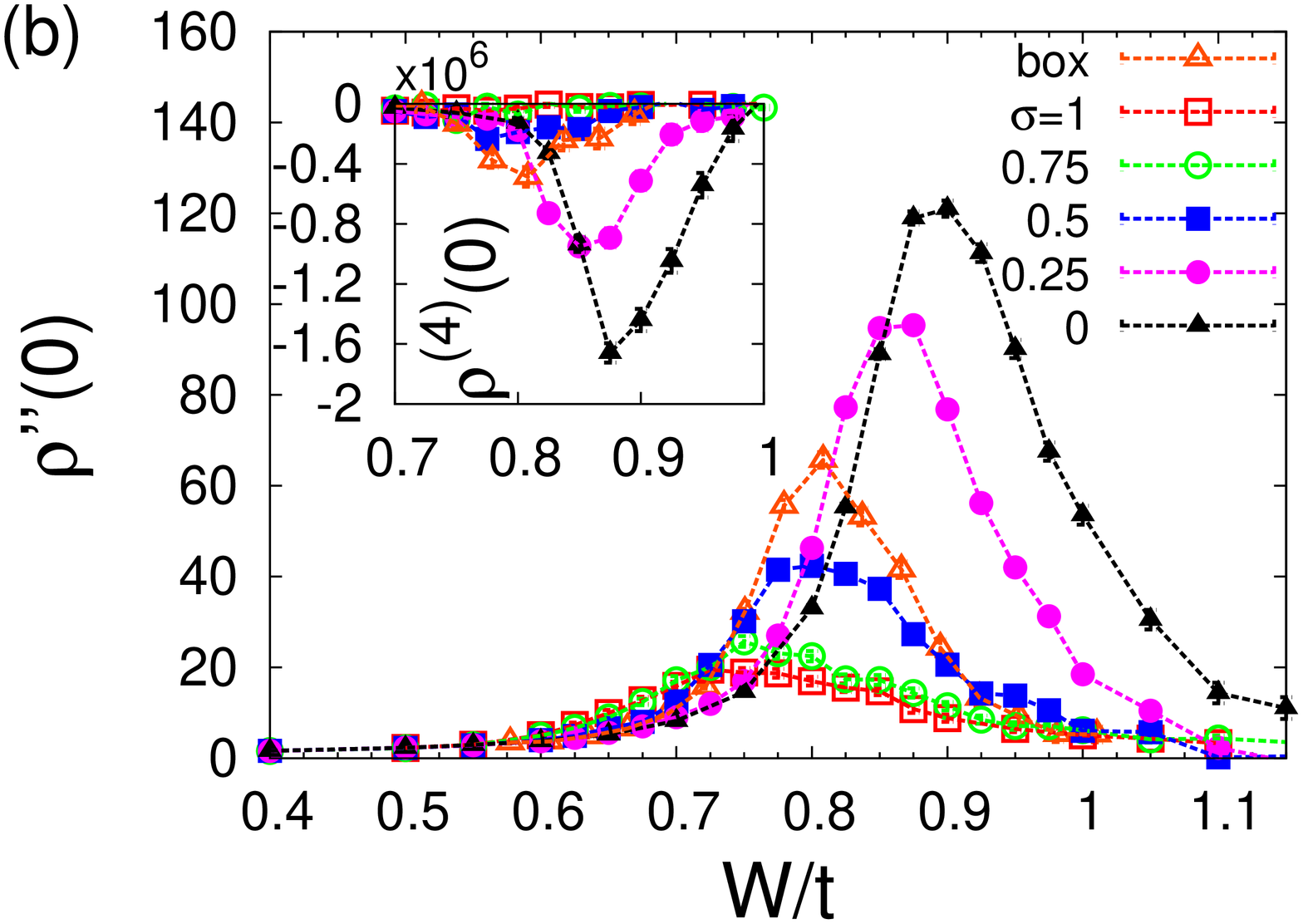}
\end{minipage}
\caption{(color online) 
$\rho''(0)$ for each $P[V]$ we consider as a function 
$W$ for a fixed expansion order $N_C=1024$ and a linear system size $L=31$. (a) The comparison of the Gaussian distribution with shifting the potential  $\tilde{V}(\br)=V(\br)-\sum_{\br}V(\br)/L^3$, and a colored Gaussian that vanishes in the long-wavelength limit. (b) Tuning the tails of the double Gaussian via $\sigma$ (and also the box distribution is shown), we find the size of the peak monotonically increases from the Gaussian case (which is small and broad) to a very large and sharp peak for binary disorder.  The inset shows $\rho^{(4)}(0)$.}
\label{fig:2}
\end{figure}

\emph{ Tuning the strength of avoidance}:
To characterize the strength of avoidance we expand the DOS (using the symmetries of the model) at low energies under the assumption that it is always analytic,
\begin{equation}
\rho(E) = \rho(0) + \frac{1}{2}\rho''(0)E^2+\frac{1}{4!}\rho^{(4)}(0)E^4+\dots,
\end{equation}
where we extract the second and fourth derivative (with respect to energy) of the DOS by directly computing them from the KPM expansion (we can also estimate them from fitting $\rho(E)$ at low $E$~\cite{supp}).  If the DOS were ever to become non-analytic $\rho''(0)$ and $\rho^{(4)}(0)$ would both diverge. Here, however, since the QCP is always rounded out, both derivatives have a peak centered very close to the location of the AQCP (see Fig.~\ref{fig:2}). Thus, we can use the \emph{magnitude of the peak} in $\rho''(0)$ and $\rho^{(4)}(0)$ to \emph{measure} the strength of the avoidance.

For the gaussian distribution the effects of rare regions are significant and the QCP is strongly avoided~\cite{Pixley-2016}. By changing $P[V]$ we make the probability to generate rare events substantially lower, which 
decreases
$E^*$ near the AQCP, and the model can thus access a larger quantum critical regime before the transition is rounded out.  As shown in Fig.~\ref{fig:2},
the size and sharpness of the peaks of $\rho''(0)$ and $\rho^{(4)}(0)$ are controlled by 
$P[V]$.
For gaussian disorder we find a very broad and weak peak, whose size is unaffected by removing the leading perturbative finite size effect or by suppressing the long wavelength components of the disorder, see Fig.~\ref{fig:2}(a).  Thus for these cases the transition is very strongly avoided.
In contrast, as shown in Fig.~\ref{fig:2}(b),
for binary disorder we find a very large and sharp peak, while the double gaussian naturally interpolates between these two,
and the box distribution falls in between 
$\sigma = 0.5$ and $0.25$.
 The peak in $\rho^{(4)}(0)$ is sharp and large ($\sim10^6$) 
 for binary disorder. We also find that the location of the AQCP (estimated from the peak location $W_p$) is tied to the strength of avoidance: for the binary case it is the largest and for gaussian it is the smallest, with a monotonic behavior between the two.  Stronger non-perturbative rare region effects destabilize the semi-metal moving the avoided critical point to smaller $W$ while making the transition more avoided.

\begin{figure}[h!]
\centering
\begin{minipage}{.225\textwidth}
\includegraphics[width=0.75\linewidth,angle=-90]{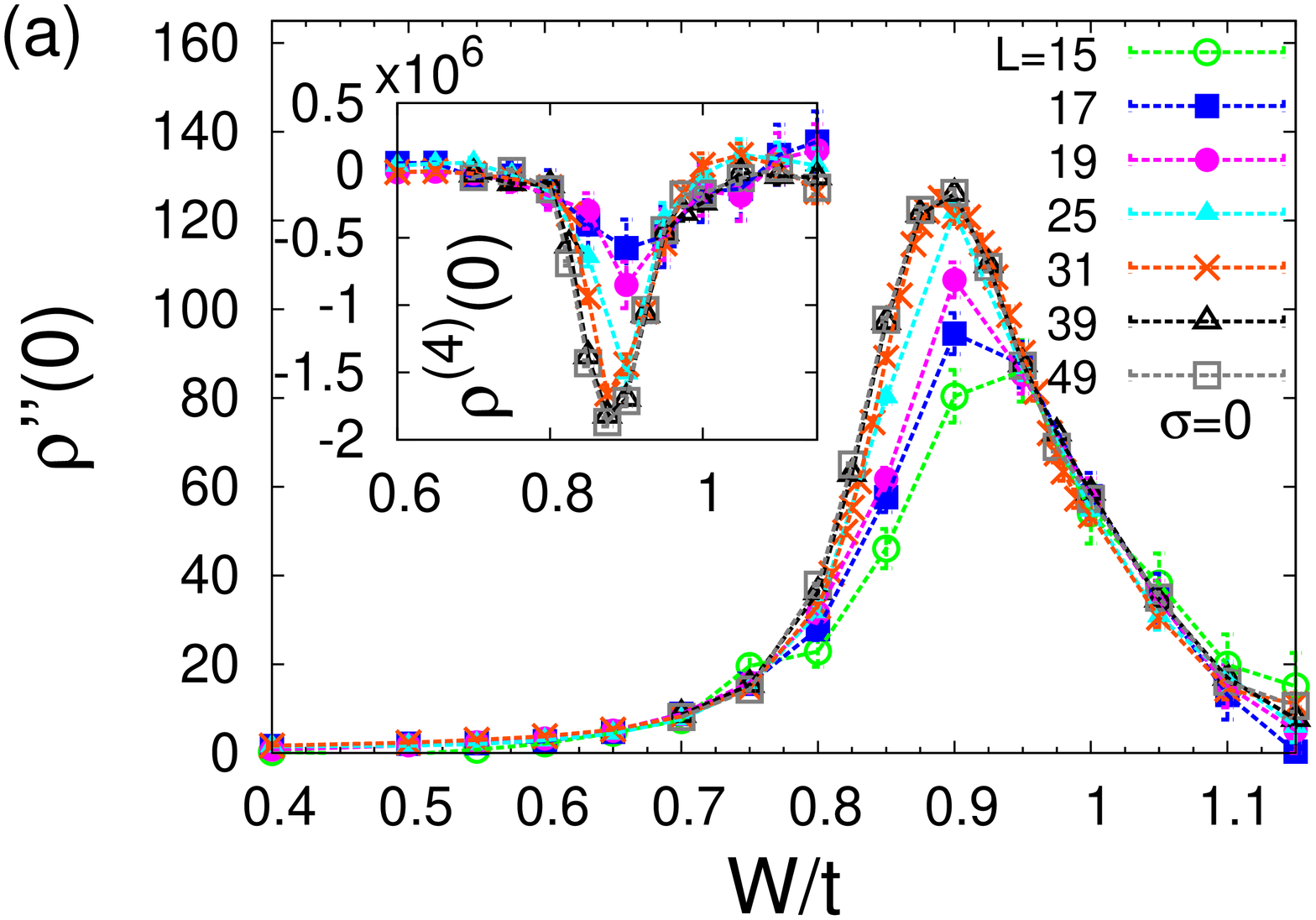}
\end{minipage}\hspace{0.5pc}
\centering
\begin{minipage}{.225\textwidth}
\includegraphics[width=0.75\linewidth,angle=-90]{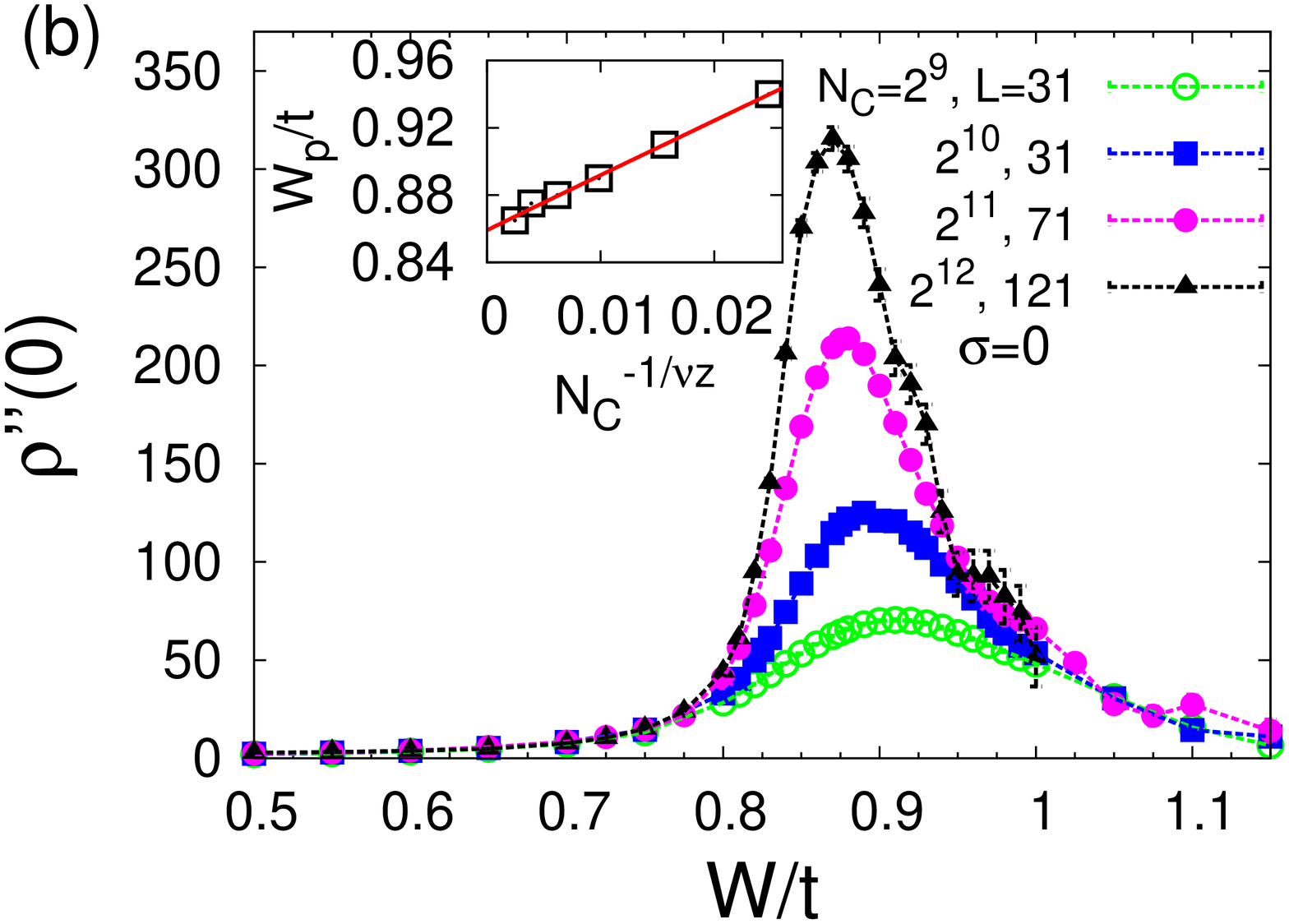}
\end{minipage}
\centering
\begin{minipage}{.225\textwidth}
\includegraphics[width=0.75\linewidth,angle=-90]{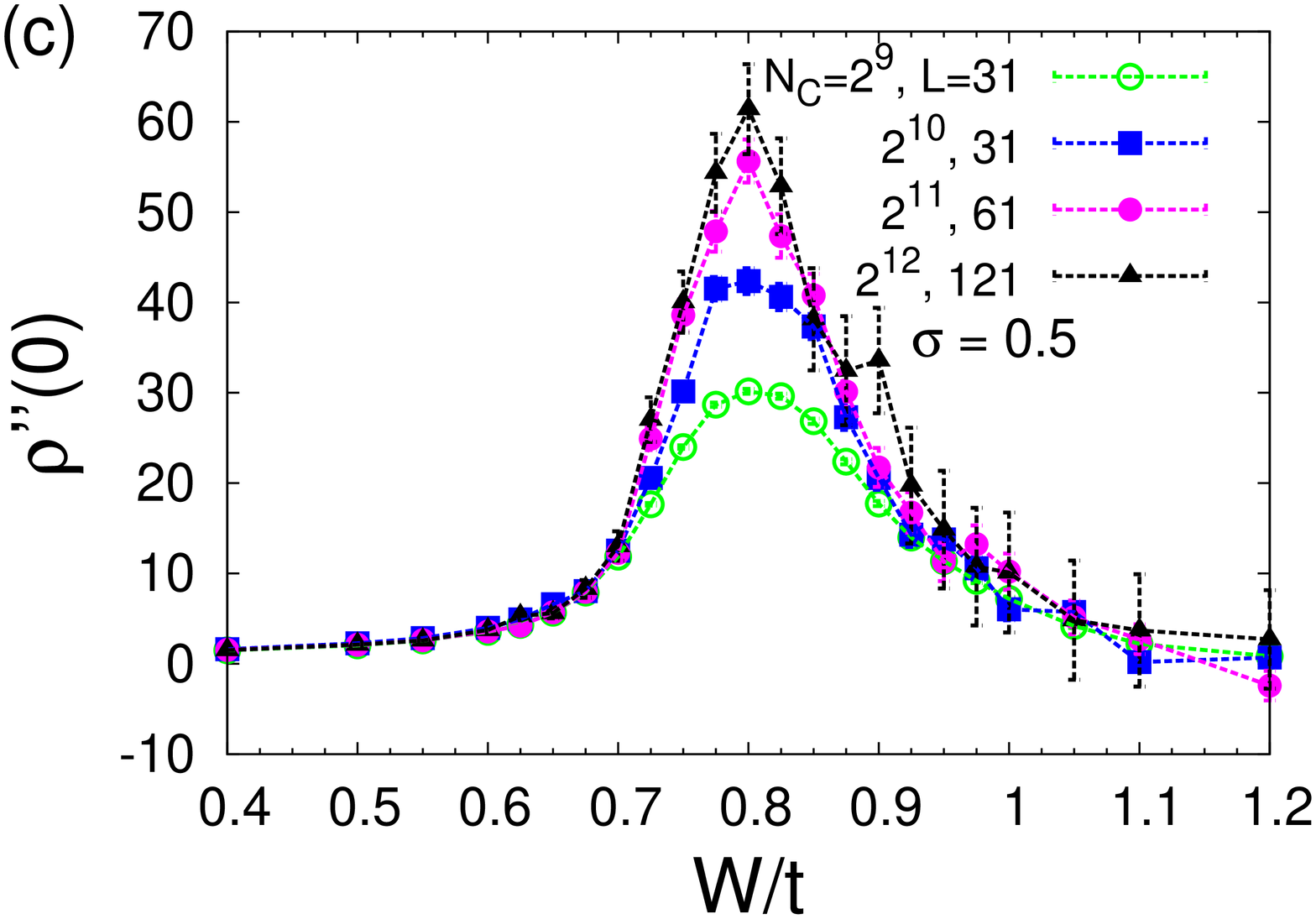}
\end{minipage}\hspace{0.5pc}
\centering
\begin{minipage}{.225\textwidth}
\includegraphics[width=0.75\linewidth,angle=-90]{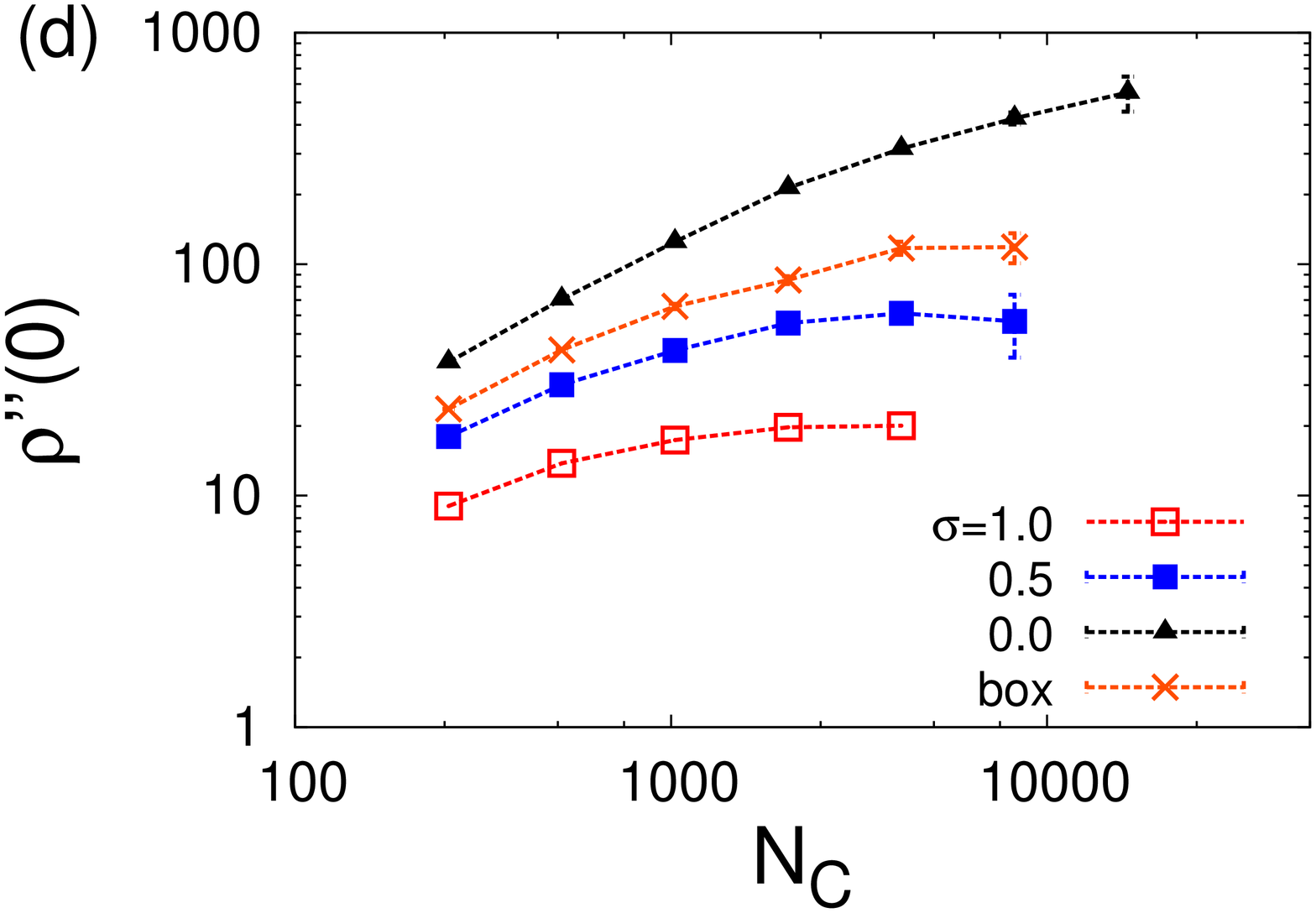}
\end{minipage}
\caption{(color online) (a) $\rho''(0)$ and (inset) $\rho^{(4)}(0)$ for binary disorder and $N_C=1024$ as a function of $W$ for various $L$.  The results are $L$-independent  
for $L\ge25$ and rounded out by finite-$L$ effects.
$\rho^{(4)}(0)$ is similar to $\rho''(0)$ but has a very large magnitude $\sim 10^6$. 
(b) $\rho''(0)$ for binary disorder as a function of $W$ for various $N_C$ ($L$ has been chosen large enough to suppress finite-$L$ rounding),  
the peak is not saturated for these $N_C$. 
 (Inset) Extrapolating the peak location $W_p$ vs $1/N_C$ using the scaling form $W_c-W_p \sim N_C^{-1/\nu z}$ 
 yielding 
 $W_c/t=0.86\pm0.01$ and $\nu z=1.5$. (c) $\rho''(0)$ as a function of $W$ for various $N_C$ and $L$ for $\sigma=0.5$. (d) The peak value of $\rho''(0)$ 
versus $N_C$ for various $P[V]$, other than $\sigma=0$ 
we can completely saturate the peak.}
\label{fig:3}
\end{figure}

In our numerical work, the transition can be rounded by finite-$L$ and by finite-$N_C$ effects, in addition to the intrinsic rounding due to non-perturbative rare region effects.  For each finite $N_C$ we go to large enough $L$ to suppress the finite-size effects~\footnote{Trying 
to 
fix
$L$ and going to large $N_C$ does not work well, because 
the data becomes very noisy due to the KPM resolving each individual eigenenergy  
as a narrow peak in the DOS.}, as illustrated in
Fig.~\ref{fig:3}(a).
However, as shown in Fig.~\ref{fig:3}(b) and (c)~\cite{supp} after we suppress the finite-size effects there still remains a strong dependence on $N_C$.  The  
broadening of the individual eigenenergies has introduced a finite length scale into the problem, which in conjunction with the non-perturbative effects is rounding out the transition.  Therefore, in order to access the regime where the transition is only rounded due to the non-perturbative effects, 
we need the results to be independent of $N_C$, which requires larger 
$N_C$ as the transition becomes less avoided.
As shown in Fig.~\ref{fig:3}(d), the peak height has a very strong dependence on $N_C$. For $\sigma=1$ the peak is saturated at 
$N_C=2048$; for $\sigma=0.5$, we find the peak is sharper, saturating at $N_C=4096$; for box disorder the peak is saturated at $N_C=8192$; and for $\sigma=0$ the peak is very sharp and perhaps still not fully saturated at a large expansion order of $N_C=16384$.
Thus for $\sigma=0$ the transition is \emph{very weakly avoided} as the evolution of the peak with $N_C$ is quite dramatic rising to a value of $\sim 550$ and $\rho^{(4)}(0)\sim10^7$~\cite{supp}.  In each case after removing all of the systematic effects of $L$ and $N_C$ the divergence of $\rho''(0)$ is always rounded out and therefore we conclude that the
non-perturbative rare region effects always induce an AQCP albeit for $\sigma=0$ this occurs at a very large length scale. Lastly we have established that the cross over energy scale $E^*=E^*(\sigma)$ decreases as $\sigma$ decreases and thus the avoidance is suppressed.

\begin{figure}[h!]
\centering
\begin{minipage}{.225\textwidth}
\includegraphics[width=0.75\linewidth,angle=-90]{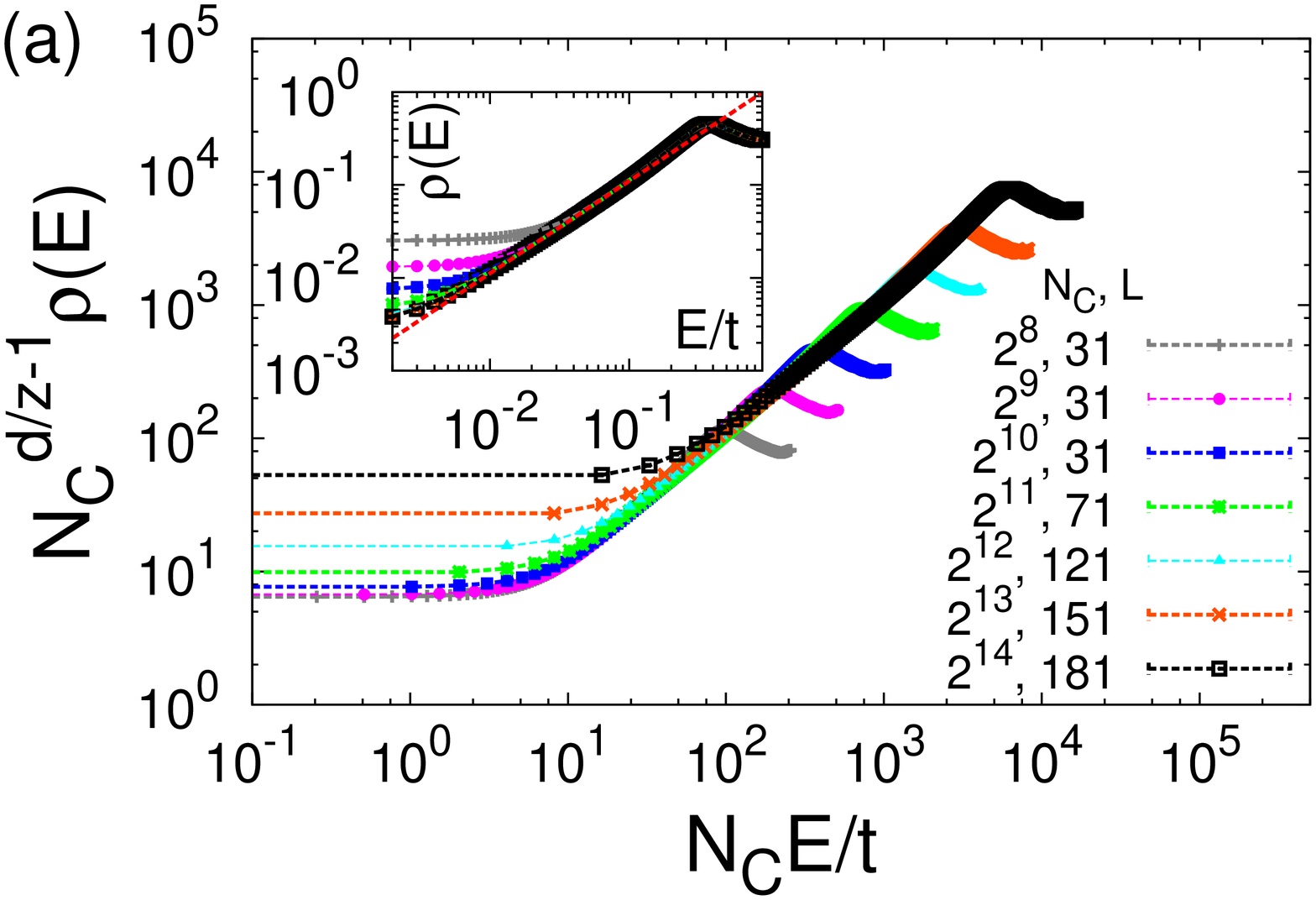}
\end{minipage}\hspace{0.5pc}
\centering
\begin{minipage}{.225\textwidth}
\includegraphics[width=0.75\linewidth,angle=-90]{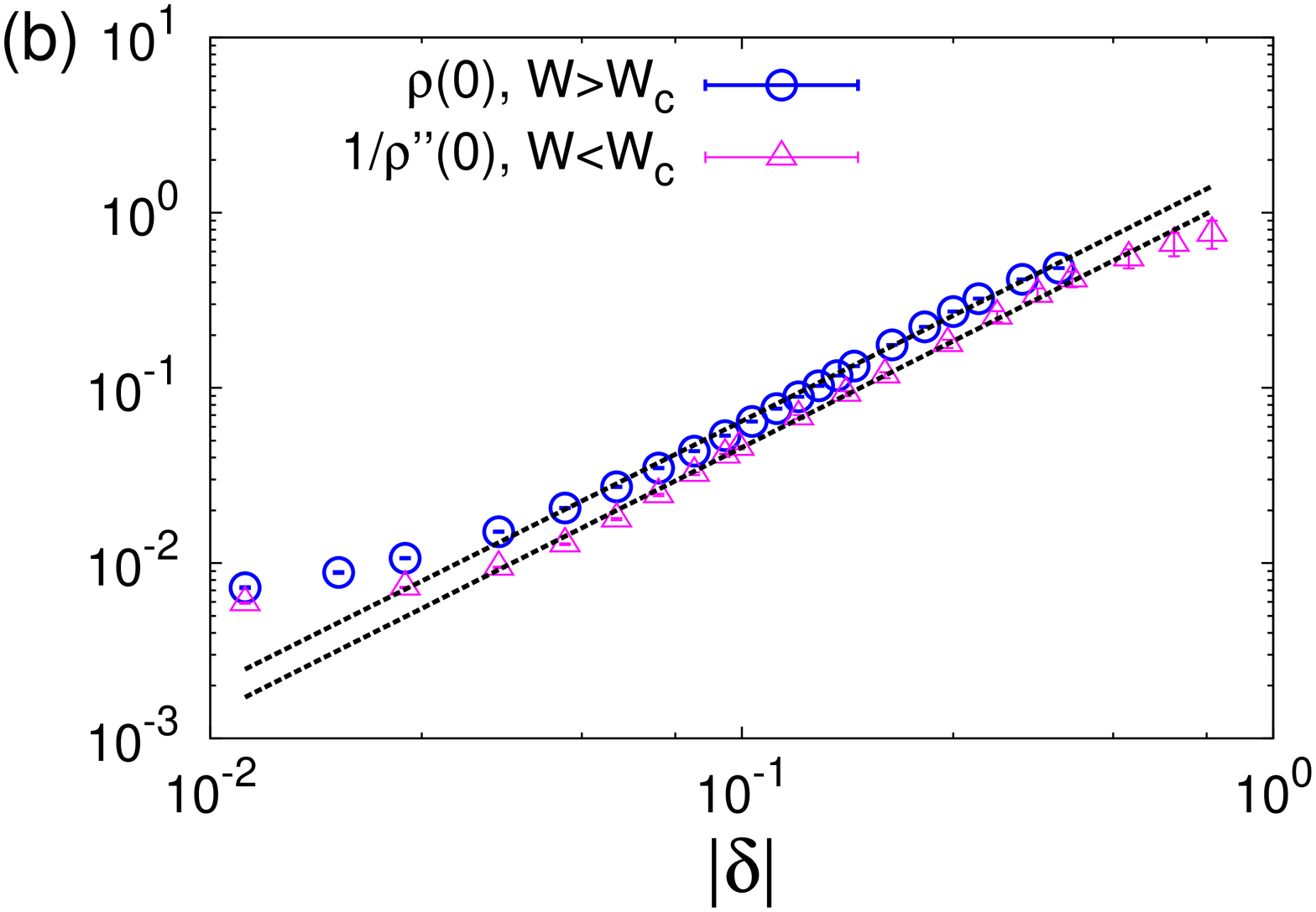}
\end{minipage}
\newline
\centering
\begin{minipage}{.225\textwidth}
\includegraphics[width=0.75\linewidth,angle=-90]{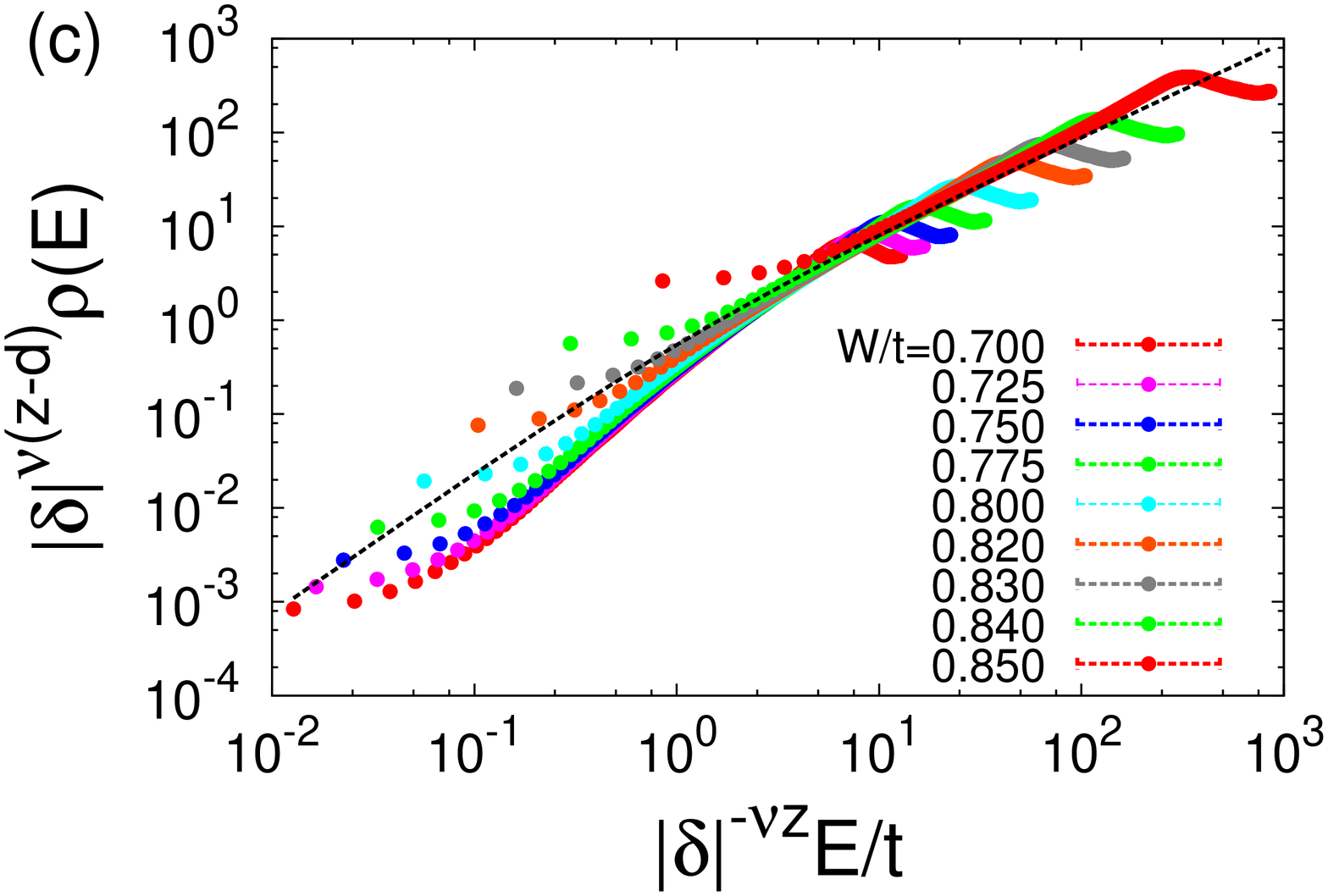}
\end{minipage}\hspace{0.5pc}
\centering
\begin{minipage}{.225\textwidth}
\includegraphics[width=0.75\linewidth,angle=-90]{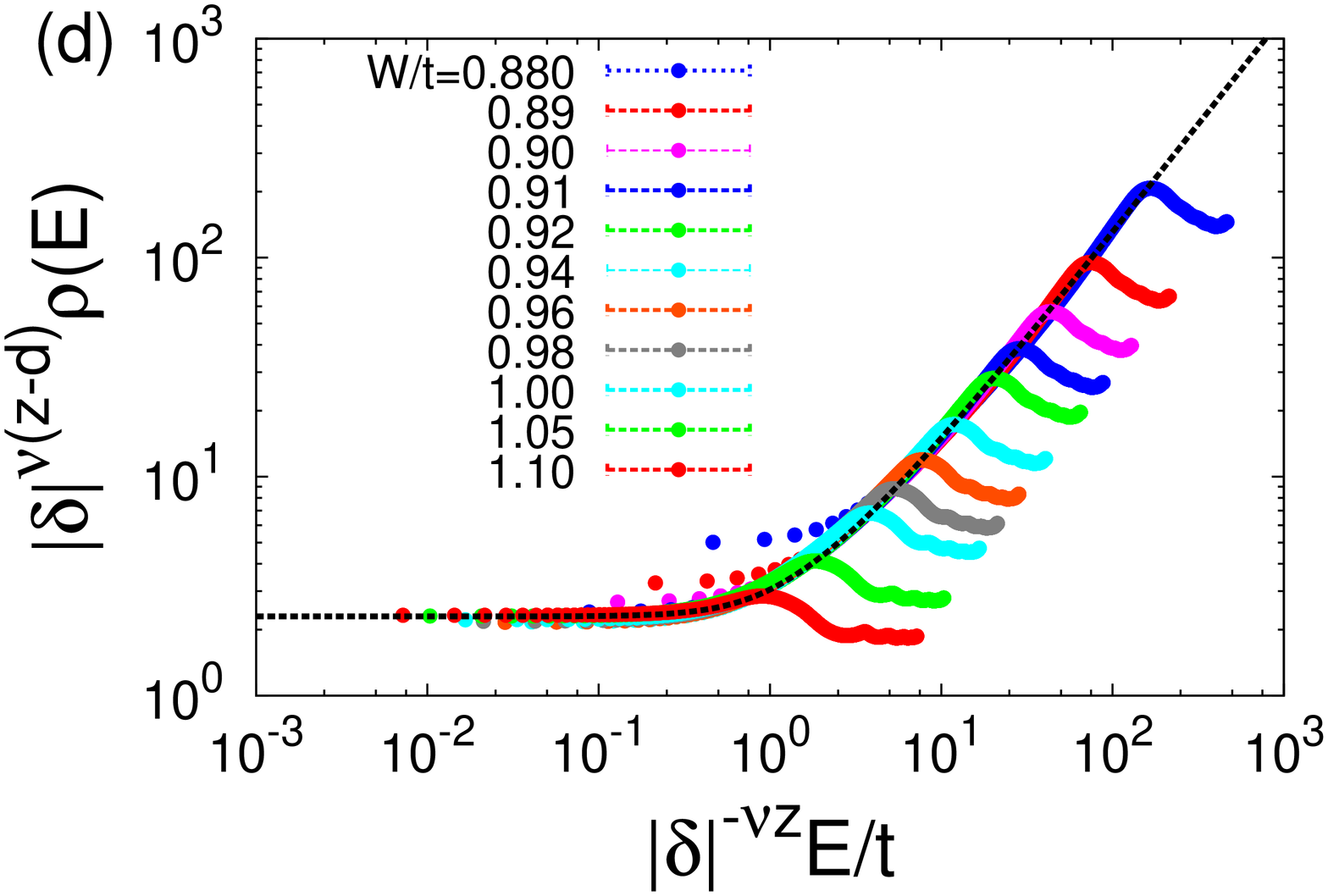}
\end{minipage}
\caption{(color online) Critical properties of the AQCP for binary disorder. (a) 
Scaling in $E$ and $N_C$
at the AQCP $W_c/t=0.86$ with an excellent scaling collapse in the QC regime for over two decades using $z=1.5$. (Inset) Dependence of $\rho(E)$ on $N_C$ at $W_c$ with a fit to the largest $N_C$ to the power law form $E^{(d/z)-1}$ yields $z=1.50\pm0.04$.  (b) Scaling in the vicinity of the AQCP in terms of $E$ and $\delta$ for $L=71$ and $N_C=2048$ for $W<W_c$ (c) and $W>W_c$ (d). Dashed lines in (c) and (d) are 
the cross over functions from the one loop RG analysis~\cite{Pixley2015disorder} (after adjusting the two bare RG scales), our data collapses onto one common curve in agreement with the cross over functions for two decades (c) and 
four decades (d).}
\label{fig:4}
\end{figure}
\emph{Properties of the AQCP}:
Since for 
$\sigma=0$ the QCP is 
\emph{very weakly avoided} we are in an excellent position to use this distribution to study quantitatively the critical properties of the AQCP, which could not be as accurately done for the other 
$P[V]$
due to the stronger avoidance.
For sufficiently weak disorder the DOS is exponentially small, i.e. for $W\ll W_c$, $ \rho(0) \sim a(W)\exp[-b(W)]$, where $a$ and $b$ depend on $P[V]$ (e.g. $b(W)\sim (t/W)^2$ for 
$\sigma=1$).  Since $\rho(0)\neq 0$ for $W\neq 0$ it cannot be used as an order parameter to estimate the location of the AQCP ($W_c$). Therefore,
we estimate $W_c$ from the location of the peak in $\rho''(0)$ as a function of $N_C$ using the scaling form $W_p-W_c\sim N_C^{-1/\nu z}$, for $\sigma=0$ we find $W_c/t=0.86 \pm $ 0.01 and $\nu z=1.5$ [see inset of Fig.~\ref{fig:3}(b)].  For this model there will be a range of $E$ and $W$ where we are far enough away from the AQCP that the avoidance is negligible, so we can study the critical behavior of the non-avoided QCP, that is actually ``hidden'' if we try to look closer.
As a function of $N_C$ at $W_c$ the scaling form in the regime where the rounding due to avoidance is negligible is
$
\rho(E,W_c,N_C) \sim N_C^{1-d/z}g(E N_C),
$
generalizing the scaling function to incorporate the rounding due to finite $N_C$ .
For a fixed $N_C$ we have $\rho(E) \sim |E|^{(d/z)-1}$ for $E^*(N_C)<E<\Lambda$ 
[see inset of Fig.~\ref{fig:4}(a)].
For $W=W_c$, $L=181$, and $N_C=16384$ (where $\rho(E)$ is independent of $L$ and $N_C$)
we fit $\rho(E)$ to a power law form (with no offset), which yields $z=1.50\pm0.04$ over a full decade.  We then collapse the data for various $N_C$'s in Fig.~\ref{fig:4}(a), which is well satisfied over two decades of $N_C E$ for $z=1.50 \pm 0.05$.  However, for $N_C > 1024$ at the lowest $|E|$ the effects of avoidance are present and the data deviate from this finite-$N_C$ scaling form, 
establishing the AQCP behavior in the 
thermodynamic limit.

At
$E=0$
we find the following power law forms~\cite{Kobayashi-2014} for the DOS and its even derivatives
\begin{eqnarray}
|\rho^{(2n)}(0)|&\sim& |\delta|^{-(z(2n+1)-d)\nu},
\end{eqnarray}
where $\delta \equiv (W-W_c)/W_c$ is the distance to the AQCP,
as shown in Fig.~\ref{fig:4}(b).  For $\rho(0)$ this holds for $W>W_c$ and we find 
$(d-z)\nu=1.51\pm0.09$, thus $\nu=1.01\pm0.06$. The divergence of $\rho''(0)$ has the same power law for $W<W_c$ and $W>W_c$, however since the statistical errors in the calculation of $\rho''(0)$ are larger for $W>W_c$ 
~\footnote{For fixed $N_C$ the error in the computed $\rho''(0)$ grows with decreasing $L$ or increasing $W$ and is even more severe for $\rho^{(4)}(0)$~\cite{supp}. 
As a result we are not able to compute $\rho^{(4)}(0)$ to high enough precision for estimating $\nu$ and $z$.} 
we only fit the power law to $\rho''(0)$ for $W<W_c$; we find $3(z-1)\nu=1.53\pm0.12$, yielding $\nu=1.02\pm0.08$.  It is interesting that the extracted numerical values of $z$ and $\nu$ are quite close to the one-loop renormalization group (RG) prediction~\cite{Goswami-2011} and deviate strongly from the two loop RG estimates~\cite{Bitan-2014,*Bitan-2016,Sergey2-2015},
which is perhaps understandable since the RG expansion parameter $(=1)$ is not small here.  
We emphasize that our estimate of the critical exponents are much more reliable than all earlier calculations in the literature which 
ignored the intrinsic rounding due to non-perturbative effects.
For $\sigma=0$ we 
find an entire decade of power law dependence 
(opposed to 
half a decade for box disorder~\cite{Pixley2015disorder}).  We stress that these data deviate from the power law closer to the AQCP because the correlation length ($\xi \sim |\delta|^{-\nu}$) is saturated by the rare region length scale ($\xi^{-z} \approx E^*$); this rounds out the transition and is neither a finite $L$ nor finite $N_C$ effect.

Far enough from the AQCP  
and at large enough $N_C$, the expected quantum critical scaling is
\begin{equation}
\rho(E,W) \sim |\delta|^{\nu(d-z)} f_{\pm}(E |\delta|^{-\nu z})~.
\label{eqn:scaling}
\end{equation}
$f_{\pm}(x)$ are scaling functions for positive and negative $\delta$.  This scaling breaks down due to the non-perturbative effects when we go too close to the avoided transition or for $W<W_c$ too close to $E=0$.  
To compare with Eq.~(\ref{eqn:scaling}), ideally we would use a large enough $N_C$ so that the rounding of the transition is purely due to the intrinsic avoidance, 
 but the required $N_C$ 
is too computationally demanding to get a complete set of such scaling data.  
Thus we use 
$N_C=2048$ and $L=71$ for 
$\sigma=0$, 
despite some of the apparent avoidance is actually due to 
$N_C$, 
as long as we use  
data far enough from the AQCP, this still allows us to study the underlying critical behavior.
The scaling collapse in $E$ and $\delta$ is quite rich: As shown in Fig.~\ref{fig:4}(c) for $W<W_c$
we find three regimes in $E\delta^{-\nu z}$. The  
DM regime $E\ll E^*$ where the data ``rolls'' off the scaling function for all these $W$, the intermediate SM regime $E^{*}<E<E_{\mathrm{SM}}$  
the data collapses for $0.7\le W/t\le 0.82$, 
and
the QC regime with $E_{\mathrm{SM}}<E<\Lambda$ where all of the data collapses onto one common curve.  
For $W>W_c$ [Fig.~\ref{fig:4}(d)] there are two scaling regimes, 
the QC regime at intermediate $\delta^{-\nu z}E/t$ and the DM regime at low energies.  
We find the collapsed data in the QC regime matches the universal cross over functions~\cite{Pixley2015disorder} obtained from a one-loop RG analysis~\cite{Goswami-2011}.

In conclusion, we have shown how to systematically control the non-perturbative effects and their associated finite (but large) length scale that always rounds out the transition.  
By making the probability to generate rare regions sufficiently low we have made the transition very weakly avoided, allowing an accurate study of the critical properties of the ``hidden'' QCP.

\acknowledgements{\emph{Acknowledgements}: We thank Igor Herbut, Olexei Motrunich, Gil Refael, Qimiao Si, Matthew Foster, Leo Radzihovsky and Sarang Gopalakrishnan for useful discussions. We also thank Pallab Goswami, Rahul Nandkishore, and Justin Wilson for various discussions and collaborations on related work.  This work is partially supported by JQI-NSF-PFC, LPS-MPO-CMTC, and Microsoft Q (JHP and SDS). DAH is supported by the Addie and Harold Broitman Membership at I.A.S.  We acknowledge the University of Maryland supercomputing resources (http://www.it.umd.edu/hpcc) made available in conducting the research reported in this paper.
}

\bibliography{DSM_RR}

\clearpage

\onecolumngrid
\setcounter{figure}{0}
\makeatletter
\renewcommand{\thefigure}{S\@arabic\c@figure}
\setcounter{equation}{0} \makeatletter
\renewcommand \theequation{S\@arabic\c@equation}
\renewcommand \thetable{S\@arabic\c@table}

\section*{Supplemental Material}

In the supplemental material we give the expression for $\rho''(0)$ within the KPM and its error with increasing $L$ and $W$. We compare the fitted value of the peak versus the direct evaluation using KPM and give more results  on $\rho''(0)$ and $\rho^{(4)}(0)$. 

The second derivative of the DOS within the KPM is obtained by analytically evaluating the second derivative of the Chebyshev expansion of the DOS with respect to energy and it is given by 
\begin{equation}
\rho''(E) = \frac{1}{\pi}g_0\mu_0\mathcal{R}_0(E) + \frac{2}{\pi} \sum_{n=1}^{N_C}g_n\mu_n \mathcal{R}_n(E),
\label{eqn:rho''}
\end{equation}
where we have introduced the functions $\mathcal{R}_j(E)$ that depend on the band width $a=(E_{\mathrm{max}}-E_{\mathrm{min}})/2$ and asymmetry $b=(E_{\mathrm{max}}+E_{\mathrm{min}})/2$. These are given by 
\begin{eqnarray}
\mathcal{R}_0(E) &=& \frac{3(E-b)^2}{(a^2-(E-b)^2)^{5/2}}+\frac{1}{(a^2-(E-b)^2)^{3/2}},
\\
\mathcal{R}_n(E) &=& \frac{a^2+2(E-b)^2}{(a^2-(E-b)^2)^{5/2}}T_n([E-b]/a) 
+ \frac{2n(E-b)}{a(a^2-(E-b)^2)^{3/2}}U_{n-1}([E-b]/a) 
\\
&+&\frac{n}{a(a^2-(E-b)^2)^{3/2}}\left[ 
(n-1)(b-E)U_{n-1}([E-b]/a)
+ anU_{n-2}([E-b]/a) \right].
\end{eqnarray}
We are denoting Chebyshev polynomials of the first and second kind as $T_n(x)$ and $U_n(x)$ respectively.
The fourth derivative can also be evaluated similarly in a straightforward but lengthy manner, however the expression is so long we do not list it here.

\begin{figure}[h!]
\begin{minipage}{.225\textwidth}
\includegraphics[width=1.0\linewidth,angle=-90]{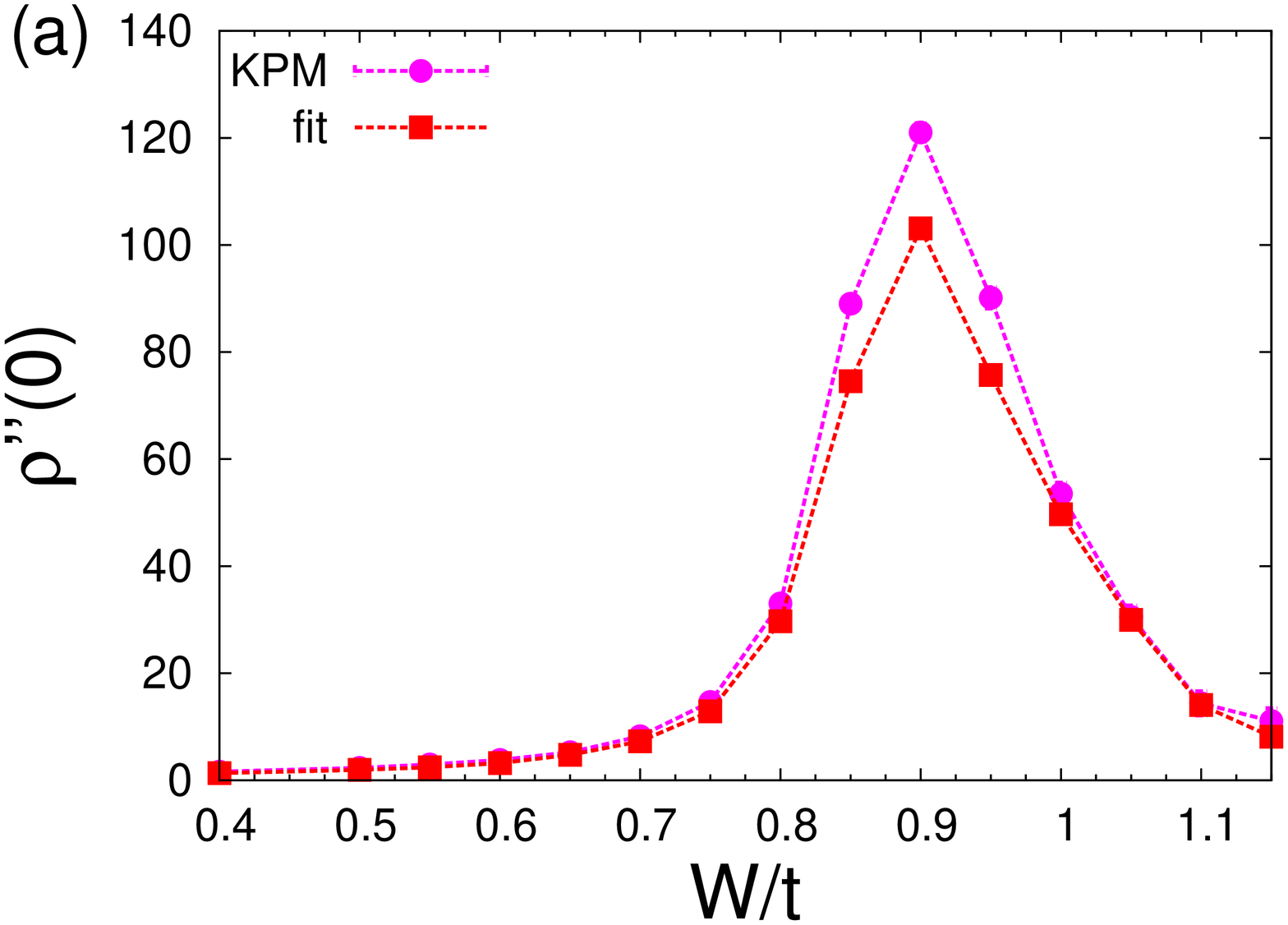}
\end{minipage}\hspace{5.pc}
\begin{minipage}{.225\textwidth}
\includegraphics[width=1.0\linewidth,angle=-90]{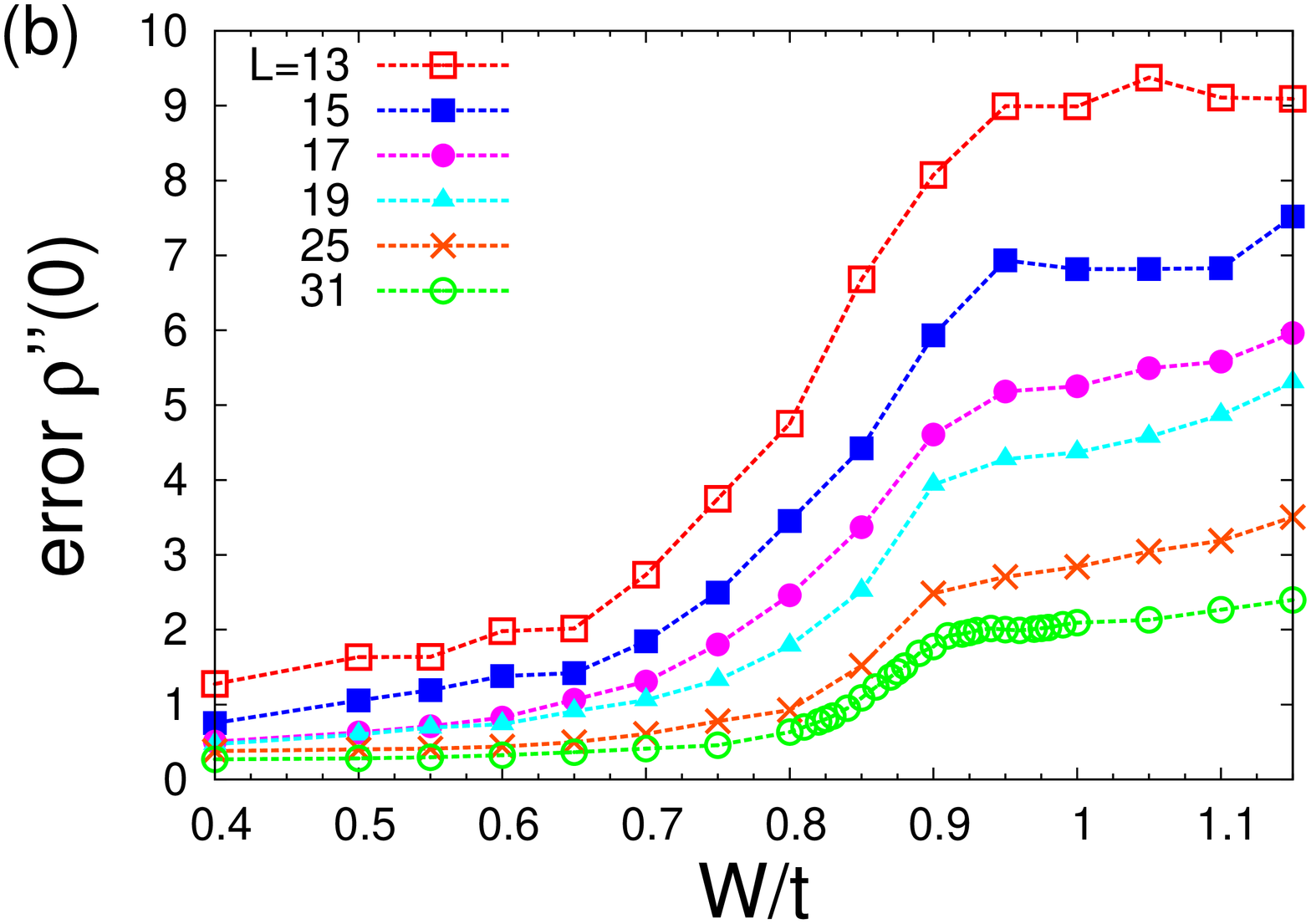}
\end{minipage}\hspace{5.pc}
\begin{minipage}{.225\textwidth}
\includegraphics[width=1.0\linewidth,angle=-90]{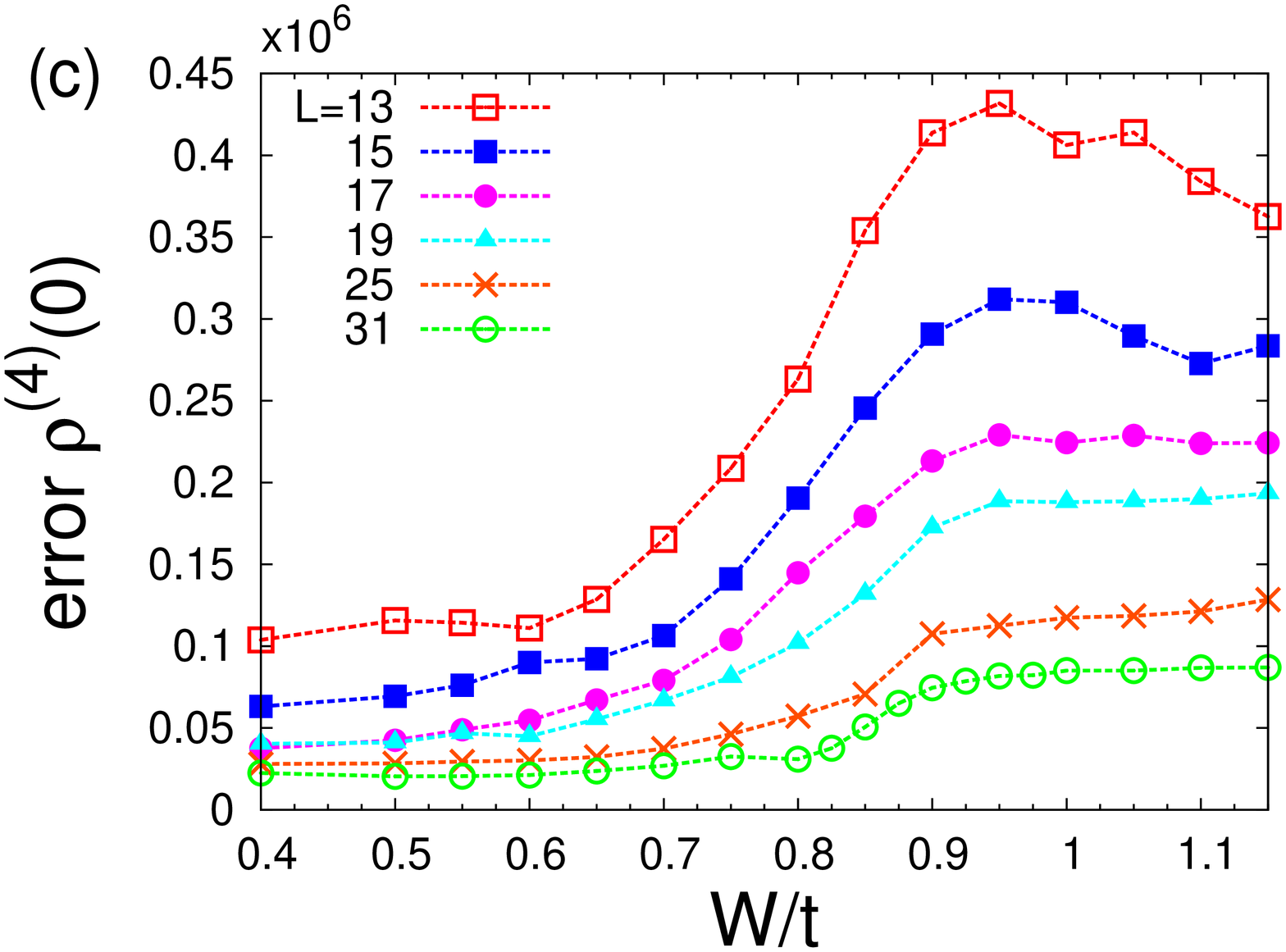}
\end{minipage}\hspace{5.pc}
\caption{(color online) Systematics of the KPM evaluated derivatives for a fixed expansion order $N_C=1024$ for binary disorder. (a) Comparison of two different ways of computing $\rho''(0)$ via fitting the low energy dependence of $\rho(E)-\rho(0)$ to $\rho''(0)E^2/2$ and directly via the KPM using equation Eq. (\ref{eqn:rho''}). Statistical error on the mean of $\rho''(0)$ (b) and $\rho^{(4)}(0)$ (c) directly computed  with the KPM. We find the error monotonically increases with increasing $W$ and decreases for increasing $L$.}
\label{fig:S1}
\end{figure}

\begin{figure}[h!]
\centering
\begin{minipage}{.3\textwidth}
\includegraphics[width=1.0\linewidth, angle=-90]{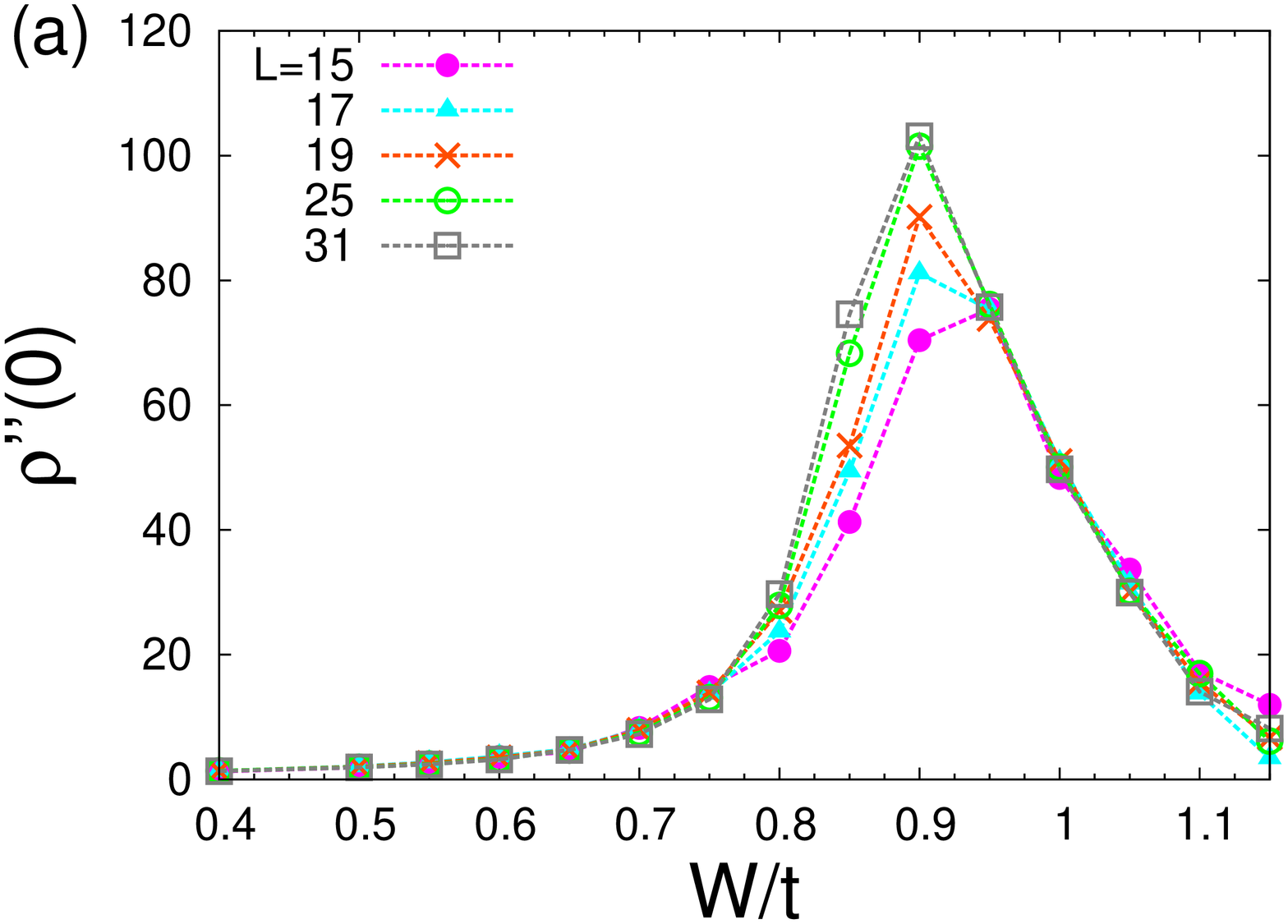}
\end{minipage}\hspace{5.5pc}
\begin{minipage}{.3\textwidth}
\includegraphics[width=1.0\linewidth, angle=-90]{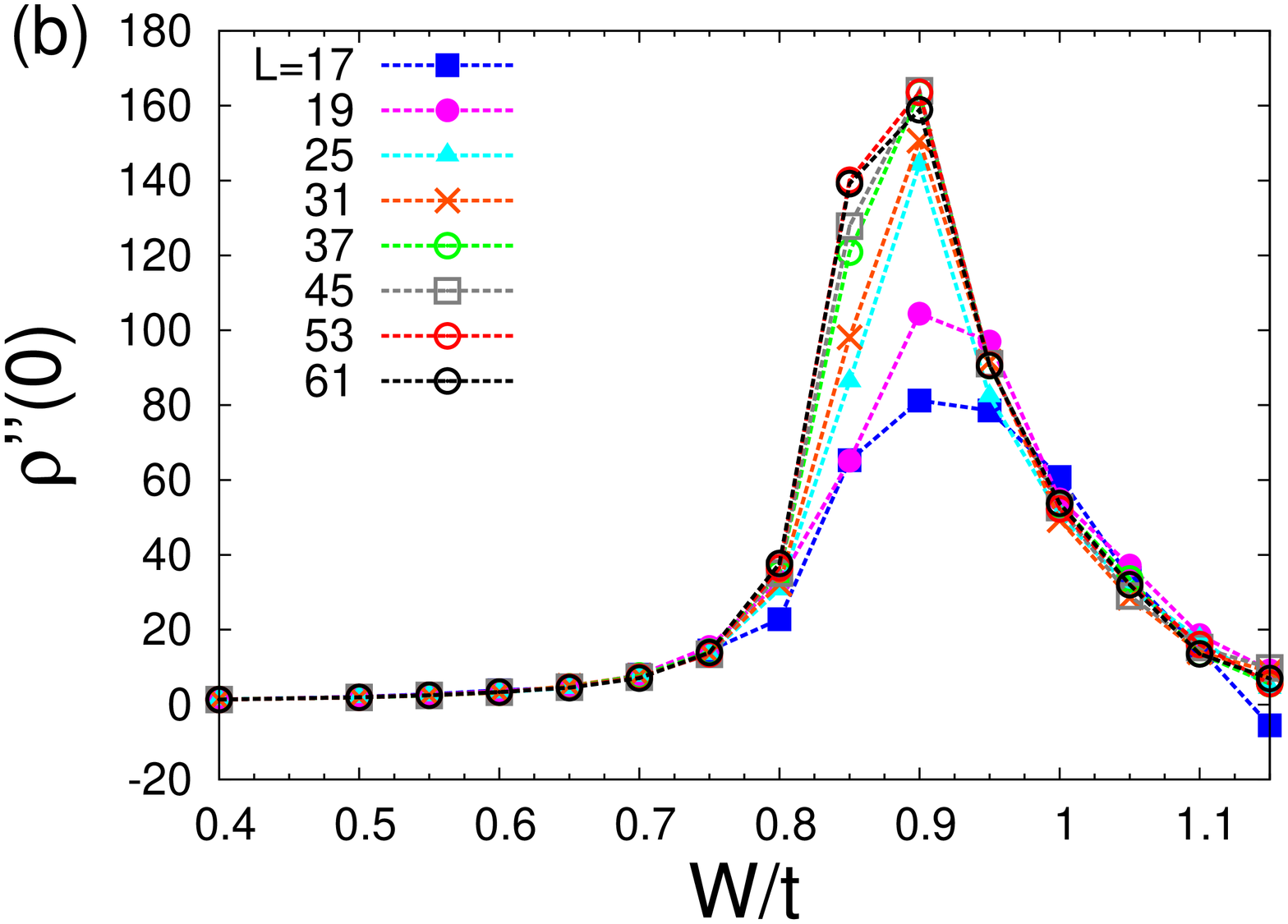}
\end{minipage}
\caption{(color online) Systematic rounding of the peak in $\rho''(0)$ (extracted from fitting the DOS) due to a finite system size $L$ for $N_C=1024$  for binary disorder (a)  and $N_C=4096$ (b). }
\label{fig:S2}
\end{figure}

We find that the fit always underestimates the size of the peak, as is natural since the fit is a more restrictive measure of the second derivative, see Fig.~\ref{fig:S1} (a). It is also not straightforward to estimate the statistical error in the fit, where as we can directly compute the error bars for $\rho''(0)$ and $\rho^{(4)}(0)$, see Figs.\ref{fig:S1} (b) and (c). We do find that the fluctuations across $W$ are less for the fitted value of $\rho''(0)$. This allows us to get a reasonably accurate estimate of small system sizes even when $N_C$ is large. This is shown in Fig.~\ref{fig:S2}(a), where we reproduce the $L$ dependence from the direct calculation in Fig. 3(a)  of the main text for $N_C=1024$, but we are also able to see the peak systematically round out with $L$ for $N_C=4096$ as shown in Fig.~\ref{fig:S2}(b). 

\begin{figure}[h!]
\centering
\begin{minipage}{.3\textwidth}
\includegraphics[width=1.0\linewidth,angle=-90]{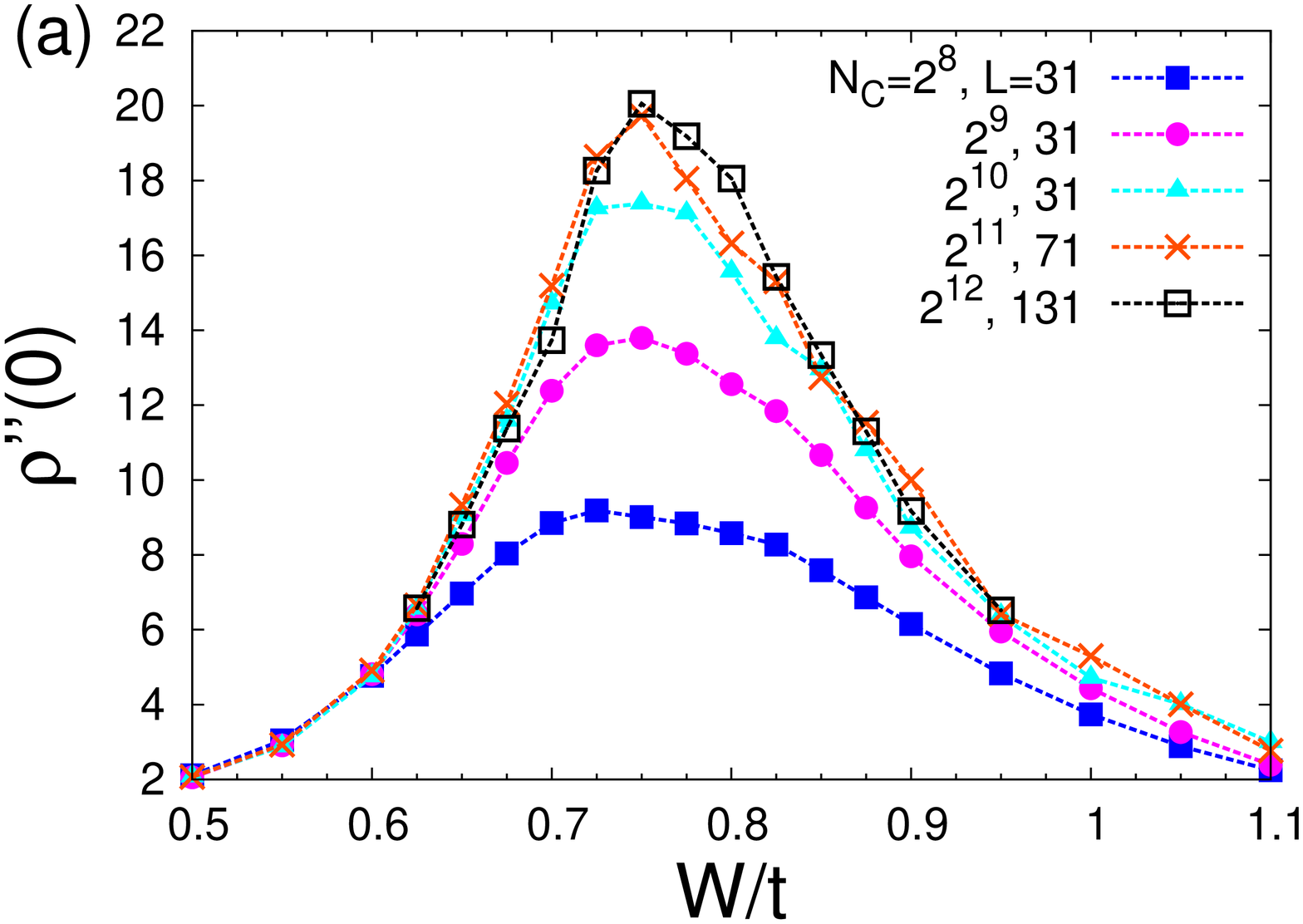}
\end{minipage}\hspace{5.5pc}
\centering
\begin{minipage}{.3\textwidth}
\includegraphics[width=1.0\linewidth,angle=-90]{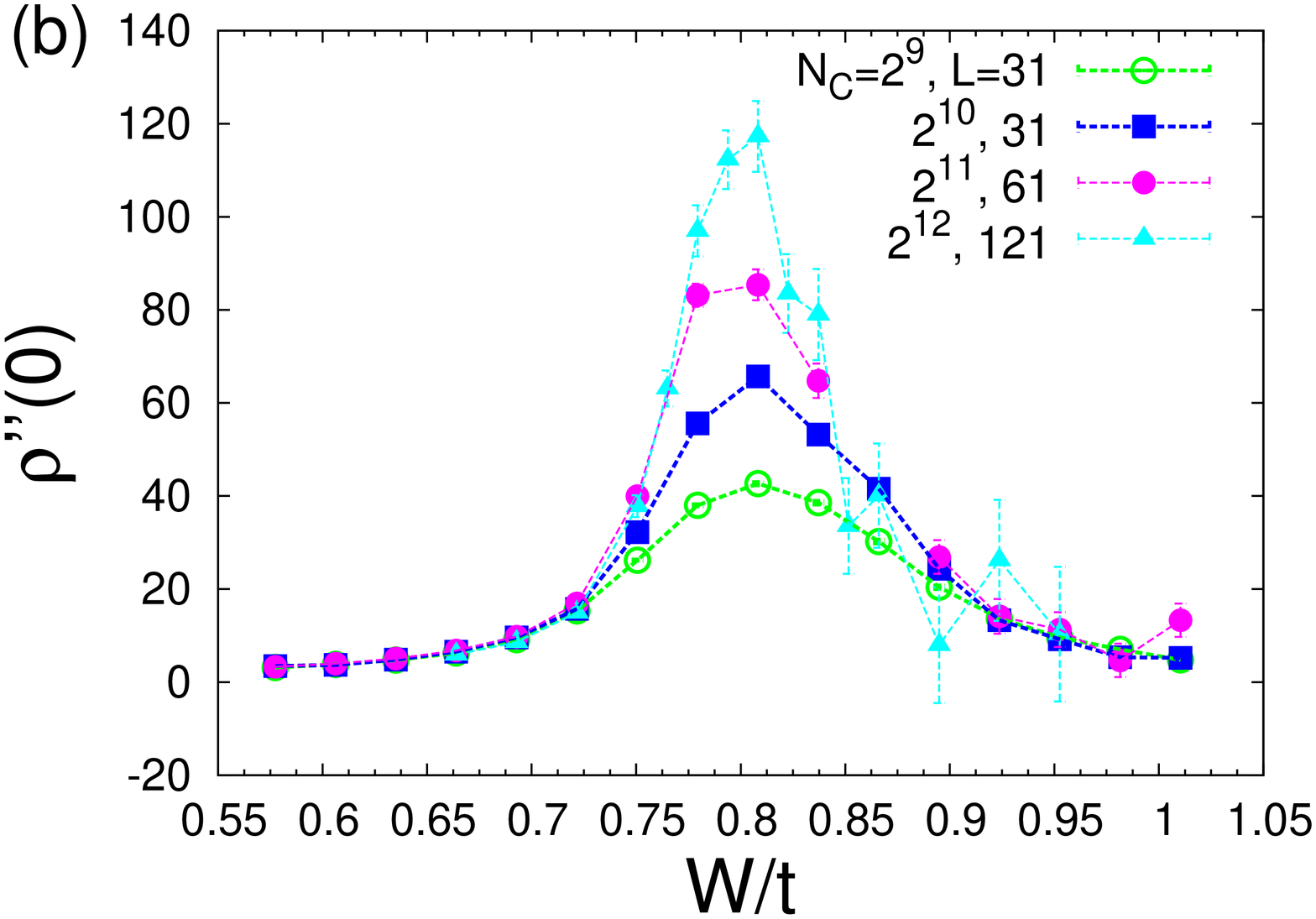}
\end{minipage}
\caption{(color online) Peak in $\rho''(0)$ for gaussian disorder with a shifted random potential extracted via fitting $\rho(E)$ (a) and box disorder from computing $\rho''(0)$ directly with KPM (b) as a function of $W$. At these values of $N_C$ and $L$ we have completely saturated the peak for Gaussian disorder but not for the box distribution. The peak for box distribution is eventually saturated at an expansion order $N_C=8192$ (see Fig. 3(d) in the main text).}
\label{fig:S3}
\end{figure}

In Fig.~\ref{fig:S3}(a) we show the results for gaussian disorder with a shifted potential $\tilde{V}(\br)=V(\br)-\sum_{\br}V(\br)/L^3$ extracting $\rho''(0)$ from the fit, which shows we can easily saturate the peak in $L$ and $N_C$. However, going to box disorder the dependence of the peak is much stronger and in Fig.~\ref{fig:S3}(b) it is not completely saturated (but it does eventually saturate as shown in Fig.3(d) of the main text).

\begin{figure}[h!]
\centering
\begin{minipage}{.3\textwidth}
\includegraphics[width=1.0\linewidth,angle=-90]{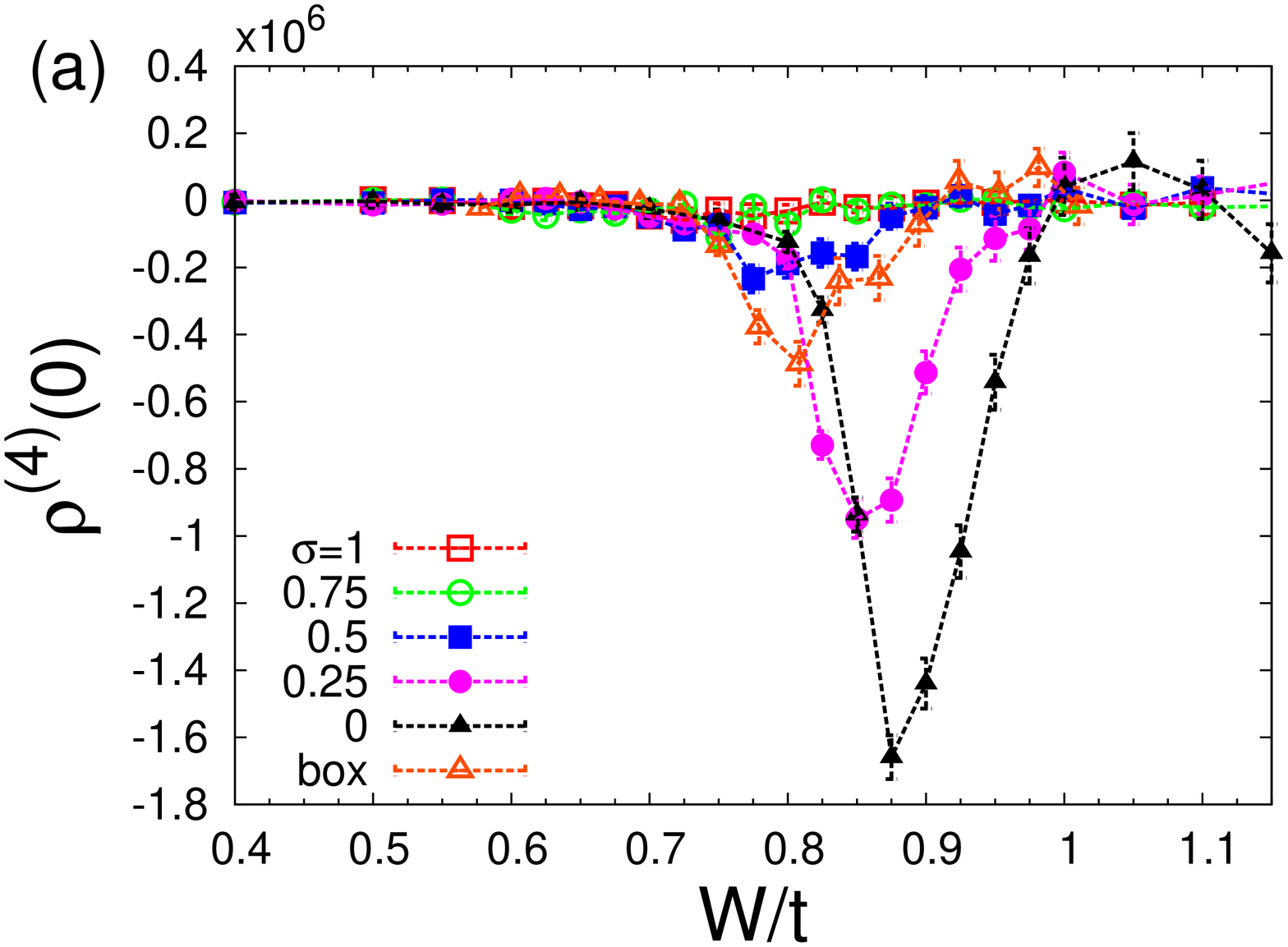}
\end{minipage}\hspace{5.5pc}
\centering
\begin{minipage}{.3\textwidth}
\includegraphics[width=1.0\linewidth,angle=-90]{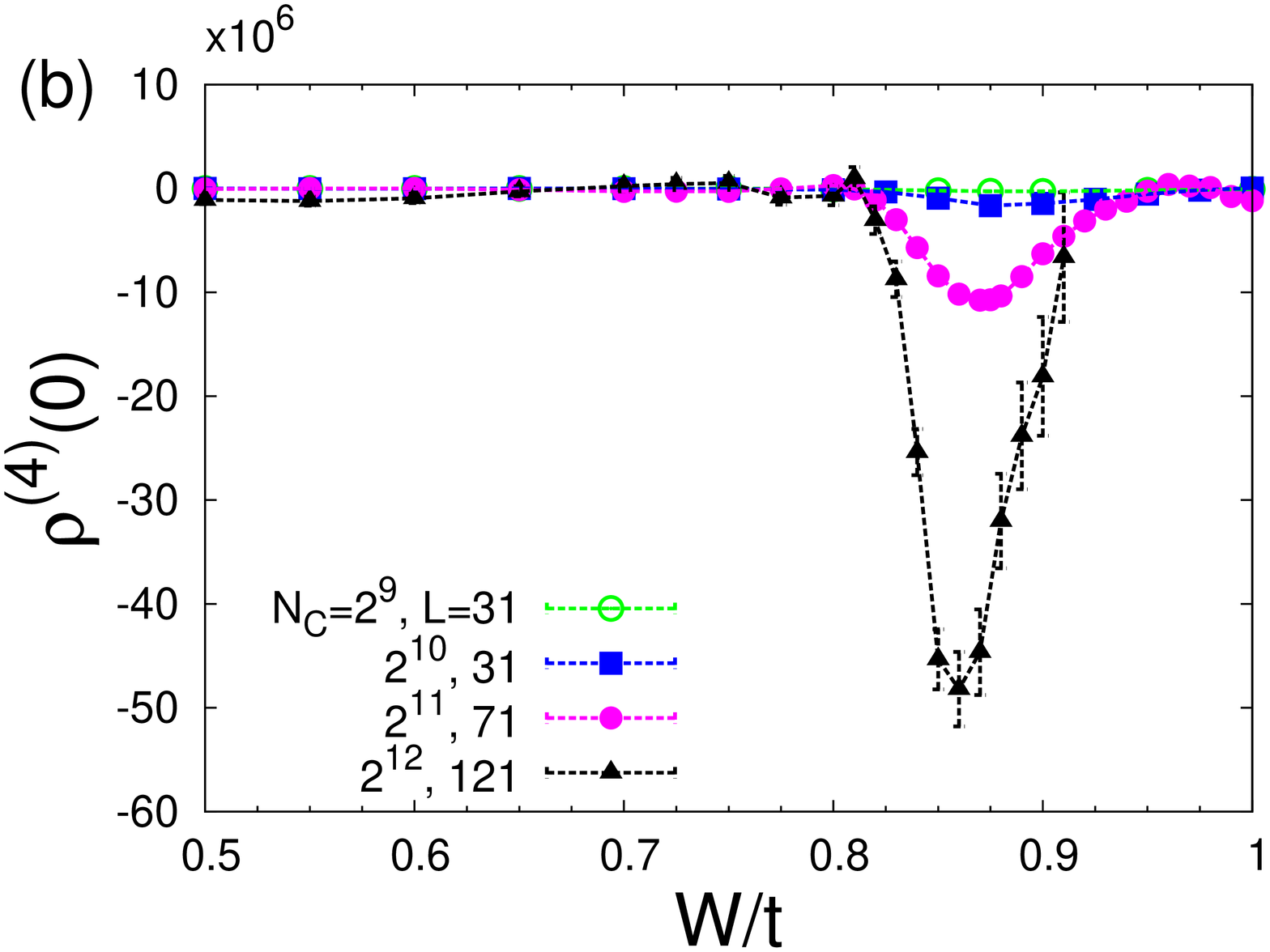}
\end{minipage}
\caption{(color online) $\rho^{(4)}(0)$ as a function of $W$ for binary disorder. The evolution of the peak in $\rho^{(4)}(0)$ as a function of $\sigma$ (this is the inset of Fig. 2(b) reproduced for clarity) displaying similar properties as $\rho''(0)$, but the fourth derivative is sharper and much larger on the order of $10^6$ for this expansion order. (b) Evolution of the peak for binary disorder as a function of $W$ for increasing $N_C$ and $L$. For $N_C=4096$ the peak in $\rho^{(4)}(0)$ is substantial on the order of $10^7$.}
\label{fig:S4}
\end{figure}

The evolution of the peak in $\rho^{(4)}(0)$ for fixed $N_C$ and various values of $\sigma$ is qualitatively similar to that of $\rho''(0)$, increasing and becoming sharper as we tune $\sigma$ from 1 to 0 but with a very large magnitude on the order of $10^6$ [see Fig.~\ref{fig:S4}(a)]. For binary disorder the peak evolves dramatically as we increase $N_C$ and $L$ becoming very sharp on the order of $10^7$ for $N_C=4096$, as shown in Fig.~\ref{fig:S4}(b).

\end{document}